\documentclass[superscriptaddress,secnumarabic,
amssymb,amsmath,nobibnotes,aps,prd,nofootinbib]{revtex4}% ,showkeys,showpacs
\usepackage{multirow}
\usepackage{graphicx}
\usepackage{epsf}
\usepackage{bm,color}
\usepackage{float}
\usepackage{amsmath}
\usepackage{amsfonts}
\usepackage{amssymb}
\usepackage{epstopdf}
\usepackage{natbib}%
\usepackage{subcaption}
\usepackage{hyperref}
\DeclareMathOperator{\sech}{sech}

\newcommand{\phiend}{\phi_{\rm end}}
\newcommand{\lambert}{{\cal LW}[\alpha, N_{\rm CMB}]}

\hypersetup{
	bookmarks=true,         % show bookmarks bar?
	unicode=false,          % non-Latin characters in Acrobat’s bookmarks
	pdftoolbar=true,        % show Acrobat’s toolbar?
	pdfmenubar=true,        % show Acrobat’s menu?
	pdffitwindow=false,     % window fit to page when opened
	pdfstartview={FitH},    % fits the width of the page to the window
	pdftitle={MHI},    % title
	pdfauthor={Author},     % author
	pdfsubject={Subject},   % subject of the document
	pdfcreator={Creator},   % creator of the document
	pdfproducer={Producer}, % producer of the document
	pdfkeywords={keyword1} {key2} {key3}, % list of keywords
	pdfnewwindow=true,      % links in new window
	colorlinks=true,       % false: boxed links; true: colored links
	linkcolor=red,          % color of internal links
	citecolor=cyan,        % color of links to bibliography
	filecolor=magenta,      % color of file links
	urlcolor=blue,           % color of external links
	linktocpage=true
}
\providecommand{\U}[1]{\protect\rule{.1in}{.1in}}
\newcommand{\be}{\begin{equation}}
	\newcommand{\ee}{\end{equation}}

\newcommand{\mincir}{\raise
	-3.truept\hbox{\rlap{\hbox{$\sim$}}\raise4.truept\hbox{$<$}\ }}
\newcommand{\magcir}{\raise
	-3.truept\hbox{\rlap{\hbox{$\sim$}}\raise4.truept\hbox{$>$}\ }}

\def\bea{\begin{eqnarray}}
	\def\eea{\end{eqnarray}}
\def\ba{\begin{array}}
	\def\ea{\end{array}}

\def\beq{\begin{equation}}
	\def\eeq{\end{equation}}

\newcommand{\eq}{Eq.\eqref}
\newcommand{\fig}{Fig.\ref}
\begin{document}
	%\title{Mukhanov Parametrization Revisited}
	\title{Mutated Hilltop Inflation in the Era of Present and Future CMB Experiments}
	\author{Barun Kumar Pal}
	\email{terminatorbarun@gmail.com}
	\affiliation{Netaji Nagar College For Women, Kolkata-700092, West Bengal, India}
	%\keywords{}
	%\pacs{}
	%%%%%%%%%%%%%%%%%%%%%%%%%%%%%%%%%%%%%%%%%%%%%%%%%%%%%%%%%%%%%%%%%%%%%%%%%%%%%%%%%%%%%%%%%%%%%%%%%%%%%%%%%%%%%%%%%%
	%%%%%%%%%%%%%%%%%%%%%%%%%%%%%%%%%%%%%%%%%%%%%%%%%%%%%%%%%%%%%%%%%%%%%%%%%%%%%%%%%%%%%%%%%%%%%%%%%%%%%%%%%%%%%%%%%%
	\begin{abstract}
	In this article we confront both large-field and small-field sectors of  mutated hilltop inflation model with the recent observational results. We begin with confrontation of predictions from mutated hilltop inflation  with the joint analysis of Planck-2018 and BICEP/Keck-2018 data. Subsequently, we extend our analysis by incorporating the ACT-DR6 data in combination with Planck-2018, BICEP/Keck-2018, and DESI-Y1 observations. In both cases, the predictions of mutated hilltop inflation show good consistency with the observational constraints. We have also forecasted the constraints on mutated hilltop inflation model from upcoming CMB experiments, LiteBIRD and Simons Observatory along with their combinations. Here also we find that the prediction from mutated hilltop inflation  are in tune with those upcoming CMB experiments. The small-field sector of mutated hilltop inflation, in principle, can probe up to $r\sim \mathcal{O}(10^{-4})$, resulting in a tensor amplitude consistent with current bounds and potentially detectable by next-generation CMB missions.  However, accommodating the high observational value of the scalar spectral index may demand relatively higher e-foldings in mutated hilltop inflation. A key appealing feature of the mutated hilltop inflation model turns out to be its ability to remain consistent with a potential non-detection of primordial gravitational waves by LiteBIRD and/or Simons Observatory.
	\end{abstract}

	%%%%%%%%%%%%%%%%%%%%%%%%%%%%%%%%%%%%%%%%%%%%%%%%%%%%%%%%%%%%%%%%%%%%%%%%%%%%%%%%%%%%
	\maketitle%% 

%%%%%%%%%%%%%%%%%%%%%%%%%%%%%%%%%%%%%%%%%%%%%%%%%%%%%%%%%%%%%%%%%%%%%%%%%%%%%%%%%%%%%%%%%%%%%%%%%%%%%%%%%%%%%%%%%%%%%%%%%%
\section{Introduction} \label{sec1}
Inflation \cite{starobinsky1978, starobinsky1980, guth1981, linde1982, starobinsky1982, linde1983, linde1983b}, a period of rapid accelerated expansion in the very early Universe, provides elegant solutions to the flatness, horizon and monopole problems of the standard Big Bang model. In addition to that it also offers a natural mechanism for the generation of curvature  perturbations that seed the large-scale structure of the Universe. During its prolonged existence, many inflationary models have been proposed \cite{lyth1999, martin2014encyclopaedia}. Scalar field models being the simplest one, which are classified  into two broad categories, \textit{large} and \textit{small-field models}, depending upon the excursion of the inflaton during observable inflation. Among the broad class of inflationary models proposed \textit{hilltop inflation} \cite{boubekeur2005, kohri2007}, which falls in the wide class of small-field models, represents a family where the scalar field rolls down from near the top of a potential, leading to a nearly scale-invariant spectrum of primordial fluctuations. An appealing aspect of hilltop inflation lies in its versatility, as a wide range of existing inflationary models can be transformed into the hilltop form through appropriate  tuning of the model parameter.

The mutated hilltop inflation (MHI henceforth) model \cite{barunmhi, barunmhip, barun2018mutated} is a variant of the conventional hilltop scenario, characterized by a potential that vanishes along with its first derivative at its absolute minimum. This feature distinguishes MHI from the standard hilltop models and ensures that inflation does not become eternal. In addition, while the usual hilltop models fall under the broad category of small-field inflation, the MHI model admits two distinct branches: one corresponding to the large-field regime and the other to the small-field regime, depending on the choice of the model parameter \cite{barun2018mutated}.  The small-field branch corresponds to scenarios where the inflaton field excursion is sub-Planckian, predicting a very small tensor-to-scalar ratio $r$, often below the detection threshold of current experiments. In contrast, the large-field branch allows super-Planckian field excursions, leading to a potentially detectable amplitude of primordial gravitational waves. This dual behaviour makes MHI an especially versatile model capable of explaining both current constraints and potential future detections. Moreover, MHI model predicts a parameter-independent scalar spectral index, which depends solely on the number of e-folds and remains nearly unaffected by the model parameter.

Nowadays, the precision achieved by present-day cosmological probes has reached remarkable levels \cite{wmap9:hinshaw2012, adame2025desi, tristram2022improved, tristram2024, aghanim2020planck, ade2018constraints, louis2025atacama,  calabrese2025atacama}, while forthcoming experiments in the likes of LiteBIRD, Simons Observatory \cite{litebird2023probing, ghigna2024litebird, SO2019} are expected to enhance this precision by a factor of two or more. Latest data released by Planck has pushed the scalar spectral index towards unity, $n_{_S}=0.969\pm 0.0035$ \cite{tristram2024}, and very recently Atacama Cosmology Telescope, ACT-DR6,  has pushed it further towards scale invariance yielding  $n_{_S}=0.9743\pm0.0034$, when combined with Planck-2018 and DESI-Y1  \cite{louis2025atacama,calabrese2025atacama,adame2025desi1, adame2025desi}. This has put age old canonical Starobinsky model \cite{starobinsky1980, Lust2024} under significant tension. At the same time, it has motivated physicists to revisit and refine existing inflationary frameworks in light of these observational advancements \cite{KalloshLindeRoest2025,dioguardi2025palatini, dioguardi2025fractional, hjquasi2025, heidarian2025alpha,ferreira2025bao, linde2025alexei, safaei2024observational}. On the other hand, the upper limit for the amplitude of primordial gravitational waves, characterized by the tensor-to-scalar ratio $r$, has been constrained to $r<0.032$ \cite{tristram2024} based on the latest combined data from Planck and BICEP/Keck observations \cite{bicep2016improved}. While forthcoming CMB space  mission LiteBIRD and ground based  Simons Observatory are promising to detect the primordial gravitational waves down to $r\sim \mathcal{O}(10^{-3})$.

With this enhanced precision of recent cosmological observations, a considerable number of theoretical models, along with their predictions, have been disfavoured or ruled out. This is an excellent juncture to test the predictions of  MHI model against observational data. The MHI model remains consistent with the latest Planck and BICEP/Keck measurements of $n_s$ and $r$. Moreover, it provides a rich phenomenological framework to be tested by upcoming CMB missions such as LiteBIRD, Simons Observatory (SO). Joint analyses combining the high-precision large-scale polarization data from LiteBIRD with the small-scale sensitivity of SO  are expected to significantly tighten constraints on $r$ as well as $n_{_S}$.  Such advancements  allow us to probe the parameter space of MHI with unprecedented precision. 
 
 In this article we have systematically confronted both large and small field sectors of MHI model with the most recent and precise observational constraints available from contemporary cosmological data sets.  
  Specifically, we have utilized the temperature and polarization measurements from the latest observational probes in the likes of Planck, ACT \cite{ade2018constraints, ade2021improved, tristram2022improved,tristram2024, calabrese2025atacama, louis2025atacama} complemented by the ground-based BICEP/Keck array \cite{bicep2016improved} which together provide stringent limits on the tensor-to-scalar ratio and the scalar spectral index. In addition to that 
 we have taken into account the large-scale structure information from DESI \cite{adame2025desi1, adame2025desi} allowing us a more comprehensive assessment of the model in light of both CMB and LSS constraints. Further we have also employed the expected sensitivities from forthcoming CMB polarization missions LiteBIRD and Simons Observatory to constrain the model parameter. The inclusion of these future forecasts allows us to assess how the parameter space of the MHI model may be constrained by the next generation of cosmological experiments.

%%%%%%%%%%%%%%%%%%%%%%%%%%%%%%%%%%%%%%%%%%%%%%%%%%%%%%%%%%%%%%%%%%%%%%%%%%5
\section{Inflationary Dynamics}
The application of  Hamilton-Jacobi formulation within the framework of inflation permits us to rewrite  the  Friedmann equations as first order second degree non-linear differential equations, where the scalar field itself is considered as the new time variable \cite{salopek1990, muslimov1990,liddle1994, kinney1997,lidsey1997, barunmhi, barunquasi,barun2018mutated, baruneos2023, barunmeos2025}, 
\bea
\left[H'(\phi)\right]^2 -\frac{3}{2\rm M_{P}^2}
H(\phi)^2&=&-\frac{1}{2\rm M_{P}^4}V(\phi)\label{hamilton}\\
\dot{\phi}&=&-2\rm M_{P}^2 H'(\phi)\label{phidot}
\eea
where prime and dot denote derivatives with respect to the scalar field $\phi$ and time respectively, and ${\rm M_{ P}}\equiv\frac{1}{\sqrt{8\pi G}}$  is the reduced Planck mass. The main advantage of this formalism is that here we only need the Hubble parameter to be specified rather than the inflaton potential. Since $H$ is a geometric quantity, unlike $V$, inflation is more naturally described in this language \cite{muslimov1990, salopek1990, lidsey1997}.
%\iffalse
Corresponding inflationary potential is then found to be   
\beq\label{potential}
V(\phi)=3\rm M_{P}^2H^2(\phi)\left[1-\frac{1}{3}\epsilon_{_H}\right]
\eeq
where $ \epsilon_{_H}$ has been defined as
\beq\label{epsilon} \epsilon_{_ H}\equiv2\rm M_{P}^2\left(\frac{H'(\phi)}{H(\phi)}
\right)^2.
\eeq 
%\fi
The acceleration equation then may be put forward as  
\beq\label{adot}
\frac{\ddot{a}}{a}=H^2(\phi)\left[1-\rm\epsilon_{_H}\right] .
\eeq
So accelerated expansion occurs whenever $\epsilon_{_{\rm H}}<1$ and ends exactly at  $\epsilon_{_{\rm H}}=1$. As a consequence,  requirement for the violation of strong energy condition is uniquely determined by $\epsilon_{_{\rm H}}<1$ only. The amount of inflation is expressed in terms of number of e-foldings, defined as
\beq\label{efol}
N(t)\equiv \ln\frac{a(t_{\rm end})}{a(t)}=\int_{t}^{t_{\rm end}} H(t)dt
\eeq 
where $t_{\rm end}$ is the time when inflation comes to an end. We have defined $N$ in such a way that at the end of inflation $N=0$ and $N$ increases as we go back in time. The observable parameters are generally evaluated when there are $50-70$ e-foldings still left before the end of inflation. Though total number of e-foldings could be much larger. During this observable period inflationary EoS may be assumed very slowly varying or even almost constant. With the help of \eq{hamilton} and \eq{epsilon},  \eq{efol} can be rewritten as a function of the scalar field as follows
\beq\label{nphi}
N(\phi)=-\frac{1}{M_P^2}\int_{\phi}^{\phi_{\rm  end}}\frac{H(\phi)}{2H'(\phi)}\ d\phi=\frac{1}{M_P}\int_{\phi_{\rm  end}}^{\phi}\frac{1}{\sqrt{2\epsilon_{_{H}}}}\ d\phi=\int_{\phi_{\rm  end}}^{\phi}\frac{1}{\epsilon_{_{H}}}\frac{H'(\phi)}{H(\phi)}\ d\phi
\eeq
where $\phi_{\rm end}$ is the value of the scalar field at the end of inflation.

It is customary to define another parameter by
\begin{equation}\label{eta}
	\eta_{_{\rm H}}=2{\rm M_{P}}^2~ \frac{H''(\phi)}{H(\phi)}. 
\end{equation}
It is worthwhile to mention here that $\epsilon_{_{\rm H}}$ and $\eta_{_{\rm H}}$ are not the usual  slow-roll parameters,  $\epsilon_{_{\rm H}}$ measures the relative contribution of the inflaton's kinetic energy to it's total energy, whereas $\eta_{_{\rm H}}$ determines the ratio of  field’s acceleration relative to the friction acting on it due to the expansion of the universe \cite{lidsey1997}. Though we do not include higher order slow-roll parameters in the present analysis, following parameters are widely used,
\bea
\zeta_{_H}^2(\phi)&\equiv& 4{\rm M_{P}}^4\ \frac{H'(\phi)H'''(\phi)}{H^2(\phi)}\\
\sigma_{_H}^3(\phi)&\equiv& 8{\rm M_{P}}^6\ \frac{H'^2(\phi)H''''(\phi)}{H^3(\phi)}.
\eea 
Slow-Roll approximation applies when $\{\epsilon_{_{\rm H}}, |\eta_{_{\rm H}}|, |\zeta_{_H}|, |\sigma_{_{\rm H}}|\}\ll 1$. Inflation goes on as long as  $\epsilon_{\rm H}<1$, even if slow-roll is broken i.e. $\{|\eta_{_{\rm H}}|, |\zeta_{_H}|, |\sigma_{_{\rm H}}|\}\gg 1$. The breakdown of slow-roll approximation  drags the inflaton towards its potential minima and end of inflation happens quickly. 

\section{Mutated Hilltop Inflation: The Model}\label{mhi}
In mutated hilltop inflation model  we deal with the following  approximate form of the Hubble parameter \cite{barunmhi, barunmhip, barun2018mutated} 
\beq\label{hubble}
H(\phi)\simeq \sqrt{\frac{V_0}{3M_{\rm P}^2}} \left[1-\sech(\alpha M_{\rm P}^{-1}\phi)\right]^{\frac{1}{2}}
\eeq 
where $V_0$ is the typical energy scale of inflation and $\alpha$ is a dimensionless parameter. For the large field value, $\alpha M_{\rm P}^{-1}\phi\gg 1$,  MHI very closely resembles $\alpha$-attractor class of 
inflationary models \cite{kallosh2013, galante2015} as we have already seen in Ref.\cite{barun2018mutated}. The end of inflation in MHI is found to be \cite{barun2018mutated}
\bea
\phi_{\rm end} &\simeq&  \frac{M_{\rm P}}{3\alpha}\sech^{-1}\left[-1+\frac{\alpha^2-6}{\alpha\left(36\alpha-\alpha^3+3\sqrt{6}\sqrt{4+22\alpha^2-\alpha^4}\right)^{1/3}}
+\frac{\left(36\alpha-\alpha^3+3\sqrt{6}\sqrt{4+22\alpha^2-\alpha^4}\right)^{1/3}}{\alpha}\right]\label{phiend}
\eea
In order to get realistic value of $\phi_{\rm end}$, we need  the argument of  $\sech$ function to be within the closed interval $[0,1]$, which gives $0<\alpha \lesssim \sqrt{11+5\sqrt{5}}$. Also, for $\alpha > \sqrt{11+5\sqrt{5}}$, $\phi_{\rm end}$ oscillates between real and complex domain. For this reason, we shall simply restrict our analysis  $\alpha\lesssim \sqrt{11+5\sqrt{5}}$. 
The number of e-foldings in MHI turns out to be  
\beq\label{efoldings}
N(\phi)\simeq\frac{1}{\alpha^2}\left[ \cosh(\alpha M_{\rm P}^{-1}\phi)- \cosh(\alpha M_{\rm P}^{-1}\phi_{\rm end}) -2 \ln\frac{\cosh(\alpha M_{\rm P}^{-1}\phi/2)}{\cosh(\alpha M_{\rm P}^{-1}\phi_{\rm end}/2) } \right].
\eeq
The above \eq{efoldings} can be analytically inverted to get the scalar field as a function of e-foldings as follows 
\bea\label{phin}\
\phi &\simeq&M_{\rm P} \alpha^{-1}\cosh^{-1}\left[-1-{\cal W}_{-1}\bigg(-\left[\cosh(\alpha M_{\rm P}^{-1}\phiend)+1\right]e^{-\alpha^2 N - 1 -\cosh(\alpha M_{\rm P}^{-1}\phiend)} \bigg)\right]\nonumber\\
%&=& \alpha^{-1}\cosh^{-1}\left[-1-{\cal W}_{-1}\bigg(-\left[{\rm M_P}\phiend^{-1}+1\right]e^{-\rm M_P^2\alpha^2 N - 1 -{\rm M_P}\phiend^{-1}} \bigg)\right]\nonumber\\
&=& M_{\rm P}\alpha^{-1}\cosh^{-1}\left({\cal LW}[\alpha,N] \right)
\eea 
where we have defined 
${\cal LW}[\alpha,N]\equiv -1-{\cal W}_{-1}\left(-\left[\cosh(\alpha M_{\rm P}^{-1}\phiend)+1\right]e^{-\alpha^2 N - 1 -\cosh(\alpha M_{\rm P}^{-1}\phiend)} \right)$ and  ${\cal W}_{-1}$ is the Lambert function. The inflation goes on along this ${\cal W}_{-1}$ branch of the Lambert function in MHI. 

Amplitude of primordial gravitational waves determines excursion of the inflaton during observable inflation, first shown in Ref.\cite{lyth1997} and known as Lyth bound, given by 
\bea\label{lythd}
\Delta\phi&=&\frac{\rm m_P}{8\sqrt{\pi}}\int_0^{N_{\rm CMB}}\sqrt{r}\ dN,
\eea
where $\rm m_P={2\sqrt{2\pi}}M_P$ is the actual Planck mass and $N_{\rm CMB}$ is the number of e-foldings still left before the end of inflation when a particular mode leaves the horizon. Large field and small-field model correspond to $\Delta\phi\geq \ {\rm m_P}$ and $\Delta\phi<  \ {\rm m_P}$ respectively. One expects to get larger tensor-to-scalar ratio, $r$, where $\Delta\phi\geq \ {\rm m_P}$ due to the higher energy scale required to explain inflationary observables. For the model under consideration we have found \cite{barun2018mutated}
\bea\label{lyth}
\Delta\phi\ {\rm m_P}^{-1}&\simeq&\alpha^{-1}\cosh^{-1}\left(\lambert\right)- \alpha^{-1}\cosh^{-1}\left({\cal LW}\left[\alpha, 0\right]\right).
\eea
As shown in \cite{barun2018mutated}, the mutated hilltop model of inflation has small excursion of the inflaton for $\alpha\geq \alpha_{\Delta\phi=1}$ and large-field excursion for  $\alpha<\alpha_{\Delta\phi=1}$, where $\alpha_{\Delta\phi=1}$ is the solution of \eq{lyth} for $\alpha$ with $\Delta\phi=1\ {\rm m_P}$. So this model is capable of addressing both the large and small-field inflationary scenario for suitable values of the model parameter.

%%%%%%%%%%%%%%%%%%%%%%%%%%%%%%%%%%%%%%%%%%%%%%%%%%%%%%%%%%%%%%%%%%%%%%%%%%%%%%%
\section{Confrontation with recent observations}
\begin{figure}
	\centering
	\begin{subfigure}{.5\textwidth}
		\centering
		\includegraphics[width=.95\linewidth,height=6cm]{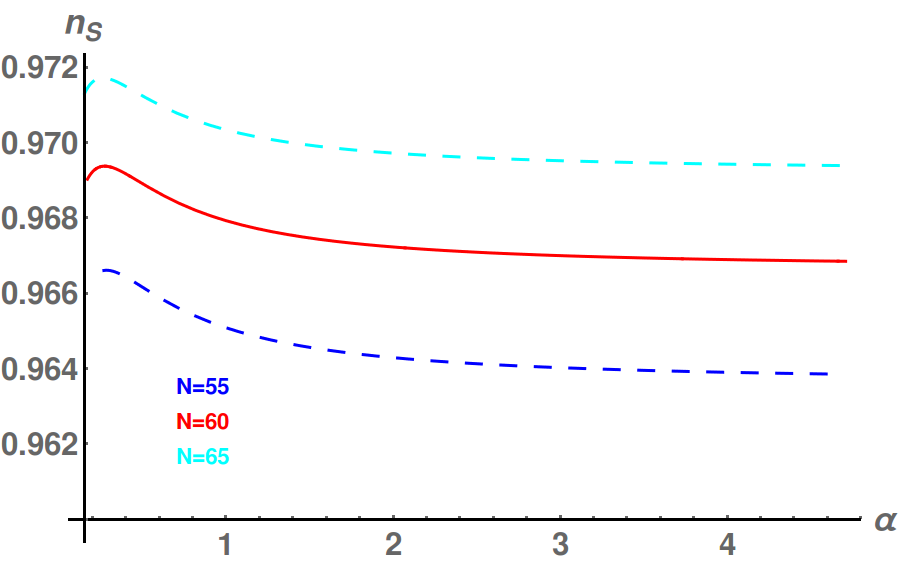}
		\caption{Scalar Spectral Index}
		%\label{fig:sub1}
	\end{subfigure}%
	\begin{subfigure}{.5\textwidth}
		\centering
		\includegraphics[width=.95\linewidth,height=6cm]{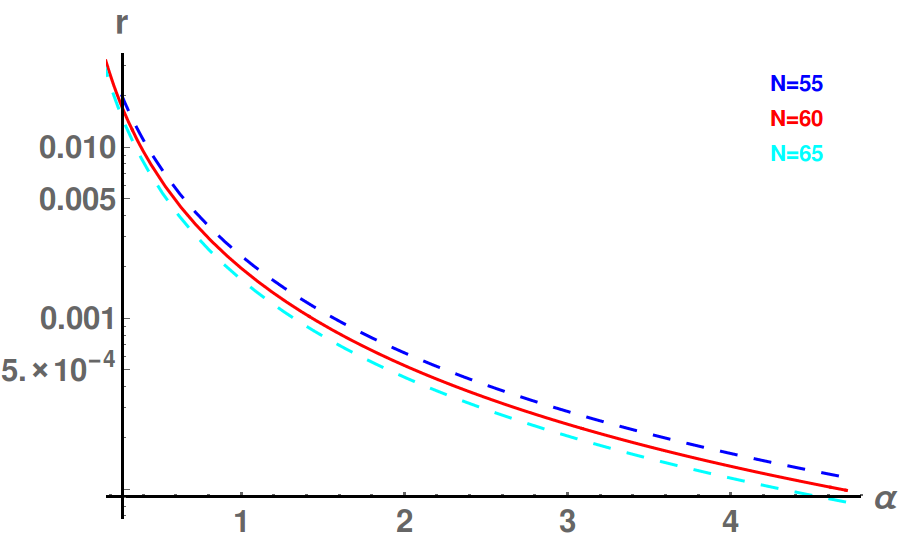}
		\caption{Tensor-to-Scalar Ratio}
		%\label{fig:sub2}
	\end{subfigure}
	\caption{Variation of the scalar spectral index (Left Panel) and tensor-to-scalar ratio (Right Panel) with the model  parameter $\alpha$ in MHI.}
	\label{fig_rns-mhi}
\end{figure}

In this work we shall systematically confront both the sectors of mutated hilltop inflation with the latest observations like ACT-DR6  in combination with Planck-2018 joint with BK18  and DESI-Y1 data \cite{calabrese2025atacama, aghanim2020planck, ade2021improved, tristram2022improved}. We have also compared inflationary observables from MHI with the joint analysis of  Planck-2018 and BK18 data. The predictions of MHI model have been tested with the expected sensitivities of upcoming CMB experiments, namely LiteBIRD~\cite{litebird2023probing} and Simons Observatory \cite{SO2019} along with their combinations. This comparison allows us to assess the extent to which these experiments will be able to constrain the inflationary parameter space predicted by the model. Before going into the observational outcomes of MHI, in the following we provide approximate formulae for the inflationary observable parameters as obtained from MHI, adopted from Ref.\cite{barun2018mutated}, 
\bea
n_{_S}&\simeq&1-\alpha^2\ \frac{2\lambert^2+3\lambert-1}{\lambert^2\left(\lambert-1\right)}\label{sr_ns}\\
r&\simeq&8\alpha^2\frac{\lambert+1}{\lambert^2\left(\lambert-1\right)}\label{sr_r}\\
\alpha_{_S}&\simeq&-\frac{\alpha^4}{2}\left[-32+30\lambert+33\cosh(2\cosh^{-1}\lambert)\right.\nonumber\\
&+&\left.\cosh(3\cosh^{-1}\lambert)\right] \lambert^{-4}(\lambert+1)^{-1}\ (\lambert-1)^{-2}.\label{sr_run}
\eea
In \fig{fig_rns-mhi} we have depicted  variation of the scalar spectral index and tensor-to-scalar ratio in mutated hilltop inflation model. The figure shows that, in MHI, $n_{_S}$ is almost independent of the model parameter, whereas $r$ varies inversely with it. Furthermore, the MHI model admits a broad range of tensor-to-scalar ratios, down to $r\sim\mathcal{O}(10^{-4})$, demonstrating its ability to encompass both small-field and large-field inflationary regimes, as evident from the figure.
%%%%%%%%%%%%%%%%%%%%%%%%%%%%%%%%%%%%%%%%%%%%%%%%%%%%%%%%%%%%%%%

\subsection{Large Field MHI}
\begin{figure}
	\centering
	\begin{subfigure}{.5\textwidth}
		\centering
		\includegraphics[width=.95\linewidth,height=6cm]{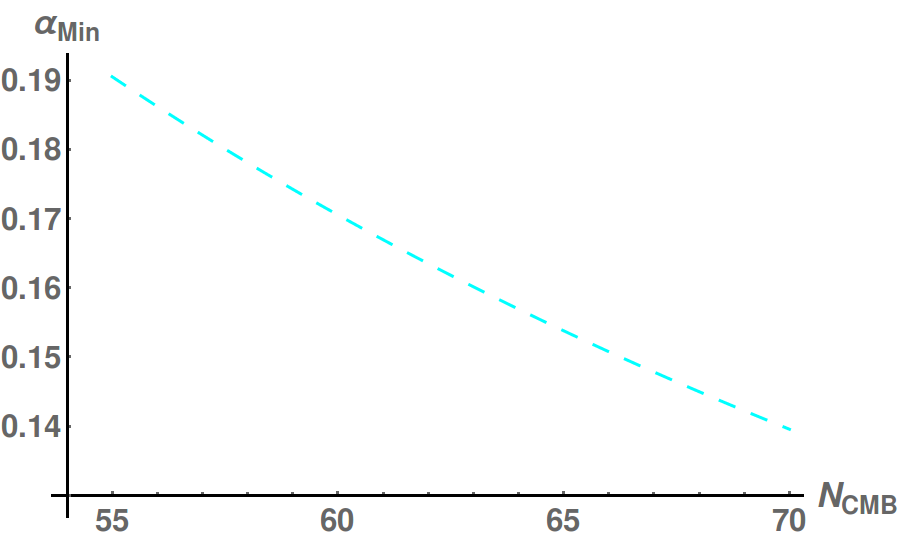}
		\caption{$\alpha_{_{\rm Min}}$}
		%\label{fig:sub1}
	\end{subfigure}%
	\begin{subfigure}{.5\textwidth}
		\centering
		\includegraphics[width=.95\linewidth,height=6cm]{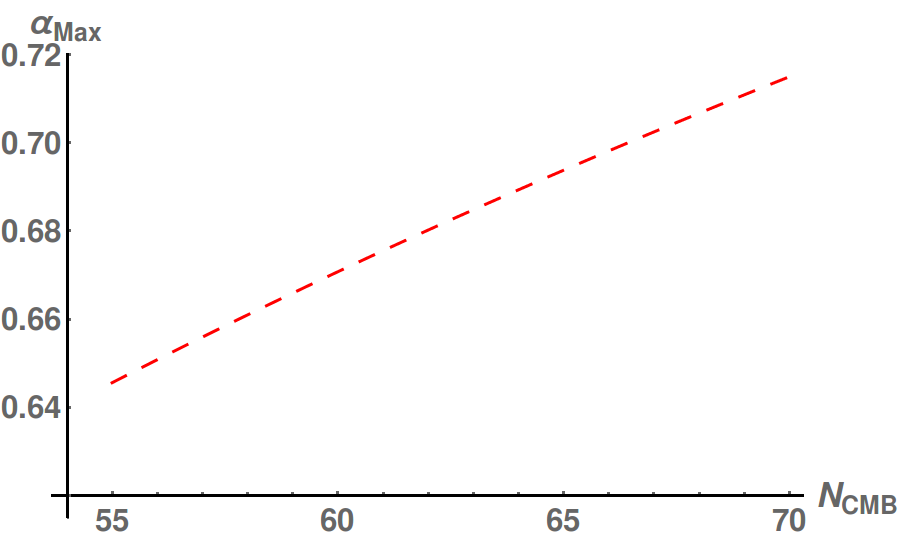}
		\caption{$\alpha_{_{\rm Max}}$}
		%\label{fig:sub2}
	\end{subfigure}
	\caption{Variation of the permeable minimum (Left Panel) and maximum (Right Panel) values of the model  parameter $\alpha$ with number of e-foldings for large-field sector  MHI. For the plot we have considered the latest constraint on primordial gravity waves   $r<0.032$ \cite{tristram2022improved} to get $\alpha_{_{\rm Min}}$ and solution of $\Delta{\phi=m_{_P}}$ for $\alpha_{_{\rm Max}}$.}
	\label{fig_alphaminmax}
\end{figure}

First we shall confront large-field sector of MHI with the latest ACT-DR6 data in combination with Planck and DESI-Y1 (P-ACT-LB henceforth) \cite{calabrese2025atacama}, Planck-2018 joint with BICEP/Keck 2018 (BK18 hence forth) \cite{tristram2022improved, ade2021improved}  along with the futuristic CMB missions, LiteBIRD \cite{litebird2023probing} and Simons Observatory \cite{SO2019}. Large field sector of MHI is supposed to produce large tensor-to-scalar ratio due to large field value required for explaining inflationary paradigm. In \fig{fig_alphaminmax} we have shown bounds on the model parameter, $\alpha$,  with the number of e-foldings. For the derivation of minimum, $\alpha_{\mbox{\footnotesize{Min}}}$, we have  utilized the  present restriction on tensor-to-scalar ratio, $r<0.032$ \cite{tristram2022improved} and maximum value, $\alpha_{\mbox{\footnotesize{Max}}}$,  is  derived by solving the equation $\Delta{\phi=1.0\ m_{_P}}$, from \eq{lyth}. This indicates that the large-field sector of the MHI model can be realized only within a relatively narrow range of the model parameter.

\begin{figure}
	\centering
	\begin{subfigure}{.5\textwidth}
		\centering
		\includegraphics[width=.95\linewidth,height=6cm]{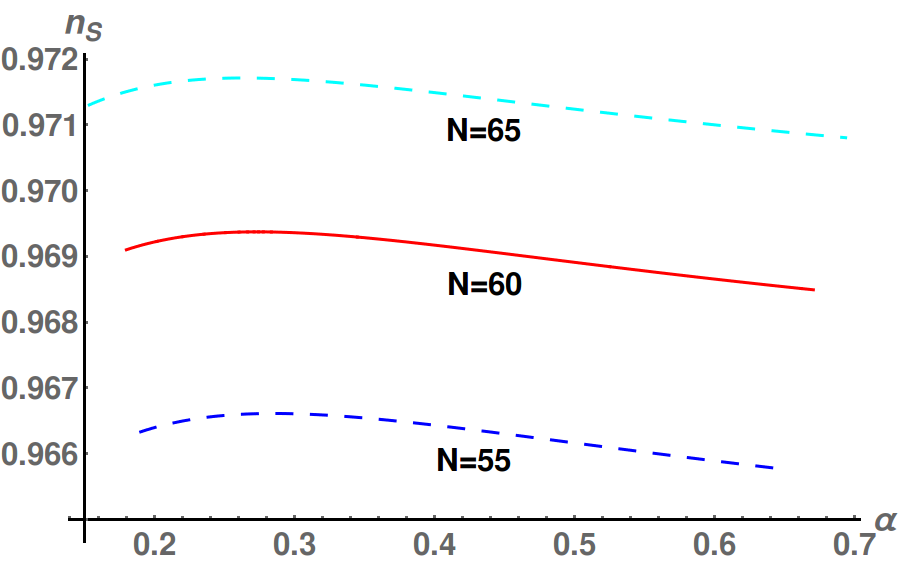}
		\caption{Scalar spectral index}
		%\label{fig:sub1}
	\end{subfigure}%
	\begin{subfigure}{.5\textwidth}
		\centering
		\includegraphics[width=.95\linewidth,height=6cm]{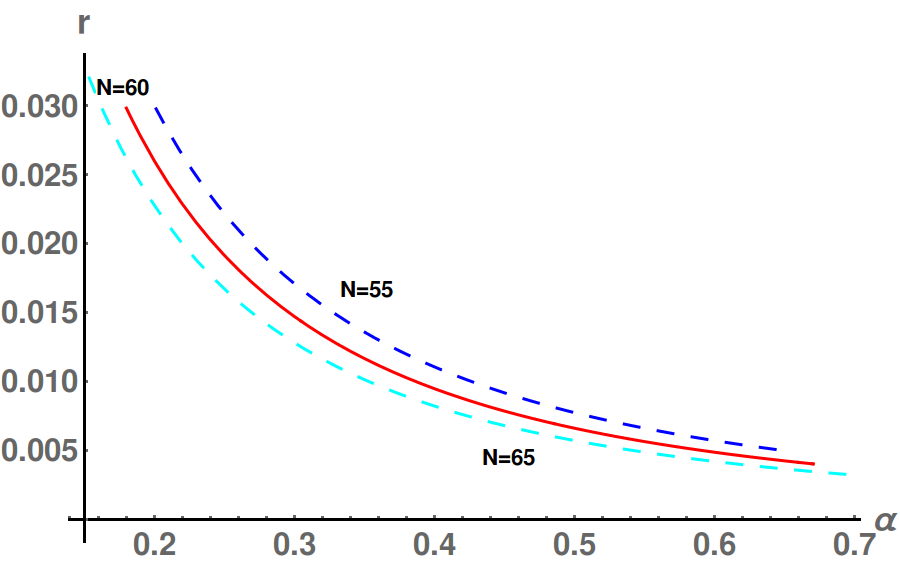}
		\caption{Tensor-to-scalar ratio}
		%\label{fig:sub2}
	\end{subfigure}
	\caption{Left Panel: The plot of scalar spectral index with the model parameter, $\alpha$, for three different values of e-foldings. The scalar spectral index exhibits negligible variation with the model parameter.   Right Panel: Plot of tensor-to-scalar ratio with the model parameter for $N=55,\ 60, \ 65$.}
	\label{fig_rns}
\end{figure}

\begin{figure}%[htb]
	\centerline{\includegraphics[width=16.cm, height=10.cm]{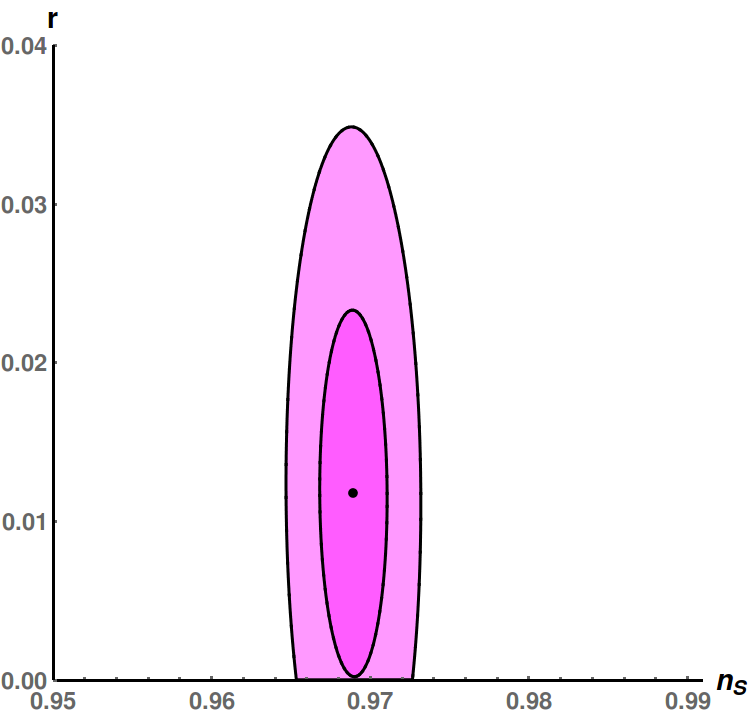}}
	\caption{\label{fig_rns-model-large}The $68\%$ and $95\%$ confidence regions in the $r-n_{_S}$ plane obtained from MHI. The contours are generated by varying the model parameter $\alpha$ within its allowed for large-field MHI  and the number of e-folds between $55$ and $65$. The black point indicates the mean prediction of the large-field MHI.}
\end{figure}

In \fig{fig_rns} we have illustrated the variations of scalar  spectral index and tensor-to-scalar ratio with the model parameter for large-field sector of MHI, for three different values of number of e-foldings. We notice that $n_{_S}$ is almost independent of $\alpha$, a noticeable signature of  MHI, while tensor-to-scalar ratio varies inversely with the model parameter. In order to visualize the range of predictions from the large-field sector of MHI model, we construct $68\%$ and $95\%$ confidence  contours in the $r$--$n_{_S}$ plane, which is generated from the statistical distribution of $n_{_S}$ and $r$ obtained through their analytical expressions in the MHI model. The resulting 1-$\sigma$ and 2-$\sigma$ ellipses, shown in Fig.\ref{fig_rns-model-large}, depicts the viable region of the model. The shaded regions represent the theoretical spread arising from variations in the parameters $\alpha$ and $N$. This plot suggests that, within the large-field MHI, the scalar spectral index and the tensor-to-scalar ratio exhibit minimal correlation, implying that variations in one have little impact on the other.

\begin{figure}%[htb]
	\centerline{\includegraphics[width=16.cm, height=10.cm]{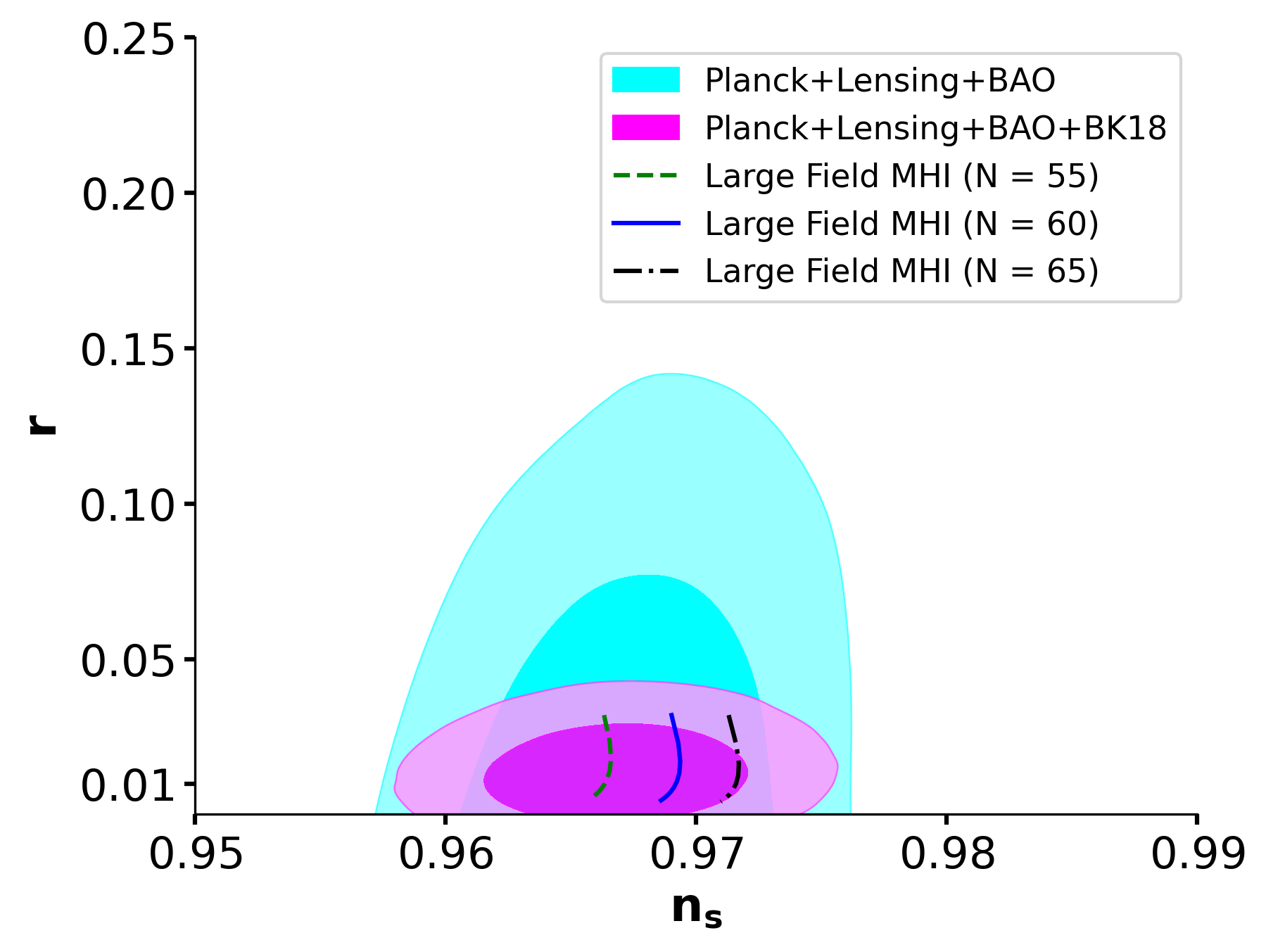}}
	\caption{\label{fig_rns-planck_large}Variation of the tensor-to-scalar ratio, $r$, with the scalar spectral index, $n_{_S}$, for three different values of e-foldings, $N = 55,\ 60,\ 65$. Black, blue, and green dashed lines correspond to predictions from the MHI (Large Field Sector) model for varying values of the model parameter and for $N = 55,\ 60,\ 65$, respectively. Marginalized $68\%$ and $95\%$ confidence regions in the plane of $r-n_{_S}$ from the Planck-2018 data \cite{planck2015inf, ade2018constraints, ade2021improved} joint with BK18 \cite{tristram2022improved}. The constraint on $r$ is driven by BICEP2/Keck (BK18) data \cite{ade2018constraints}, while the constraint on $n_{_S}$ is obtained from Planck-2018 data}
\end{figure}

\begin{figure}%[htb]
	\centerline{\includegraphics[width=16.cm, height=10.cm]{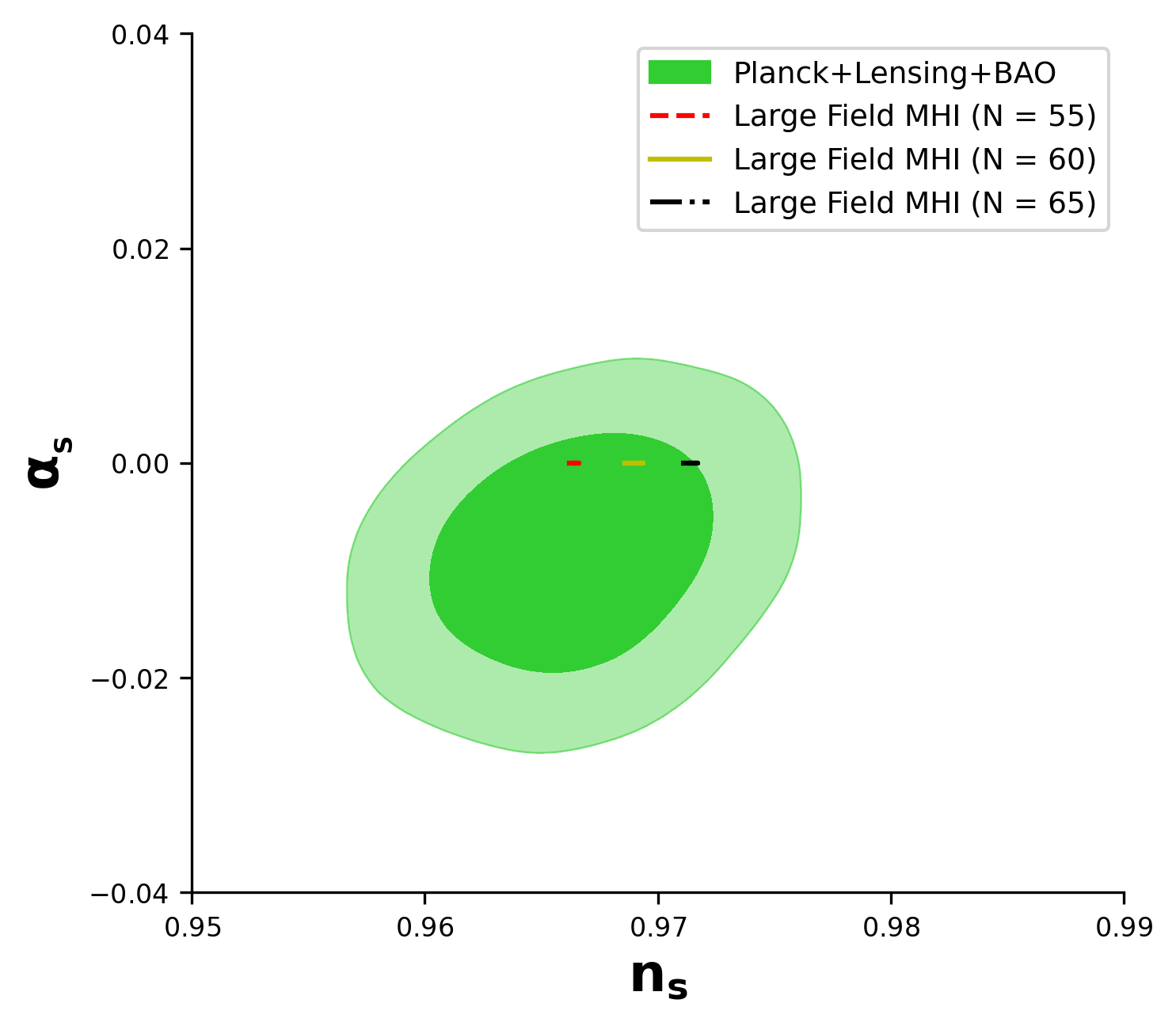}}
	\caption{\label{fig_running-planck_large}Variation of running of scalar spectral index, $\alpha_{_S}$, with the spectral index, $n_{_S}$, for three different values of e-foldings, $N = 55,\ 60,\ 65$. Black, blue, and green dashed lines correspond to predictions from the MHI (Large Field Sector) model for varying values of the model parameter and for $N = 55,\ 60,\ 65$, respectively. Marginalized $68\%$ and $95\%$ confidence regions in the plane of $\alpha_{_S}-n_{_S}$ from the Planck-2018 data with Lensing and BAO\cite{planck2015inf, ade2018constraints, ade2021improved} }
\end{figure}

In \fig{fig_rns-planck_large} we have shown prediction from the large-field MHI   in  the plane of $r$ and  $n_{_S}$, for three different values of e-foldings, $N = 55,\ 60,\ 65$, obtained by varying the model parameter within its allowed region. The shaded contours correspond to the marginalized $68\%$ and $95\%$ confidence regions from the Planck-2018 in combination with Lensing and BAO\cite{ade2018constraints, aghanim2020planck, ade2021improved} data with (Small Magenta coloured area) and without (Large Cyan coloured area) BK18 data \cite{tristram2022improved}. 
In \fig{fig_running-planck_large} we have illustrated running of scalar spectral index versus spectral index in the $\alpha_{_S}-n_{_S}$ plane, for three different values of e-foldings, $N = 55,\ 60,\ 65$ represented by red, yellow and black lines respectively. These curves correspond to the predictions from MHI (Large Field Sector) model for varying values of the model parameter. Shaded contour represents  marginalized $68\%$ and $95\%$ confidence regions in the plane of $\alpha_{_S}-n_{_S}$ from the Planck-2018 data \cite{planck2015inf, ade2018constraints, ade2021improved}. We see that large-field sector of MHI provides an excellent match with the observations for a wide range of model parameter.  
It is transparent from the figures that prediction of  large-field MHI are in tune with Planck-2018 data.

\begin{figure}%[htb]
	\centerline{\includegraphics[width=15.cm, height=9.cm]{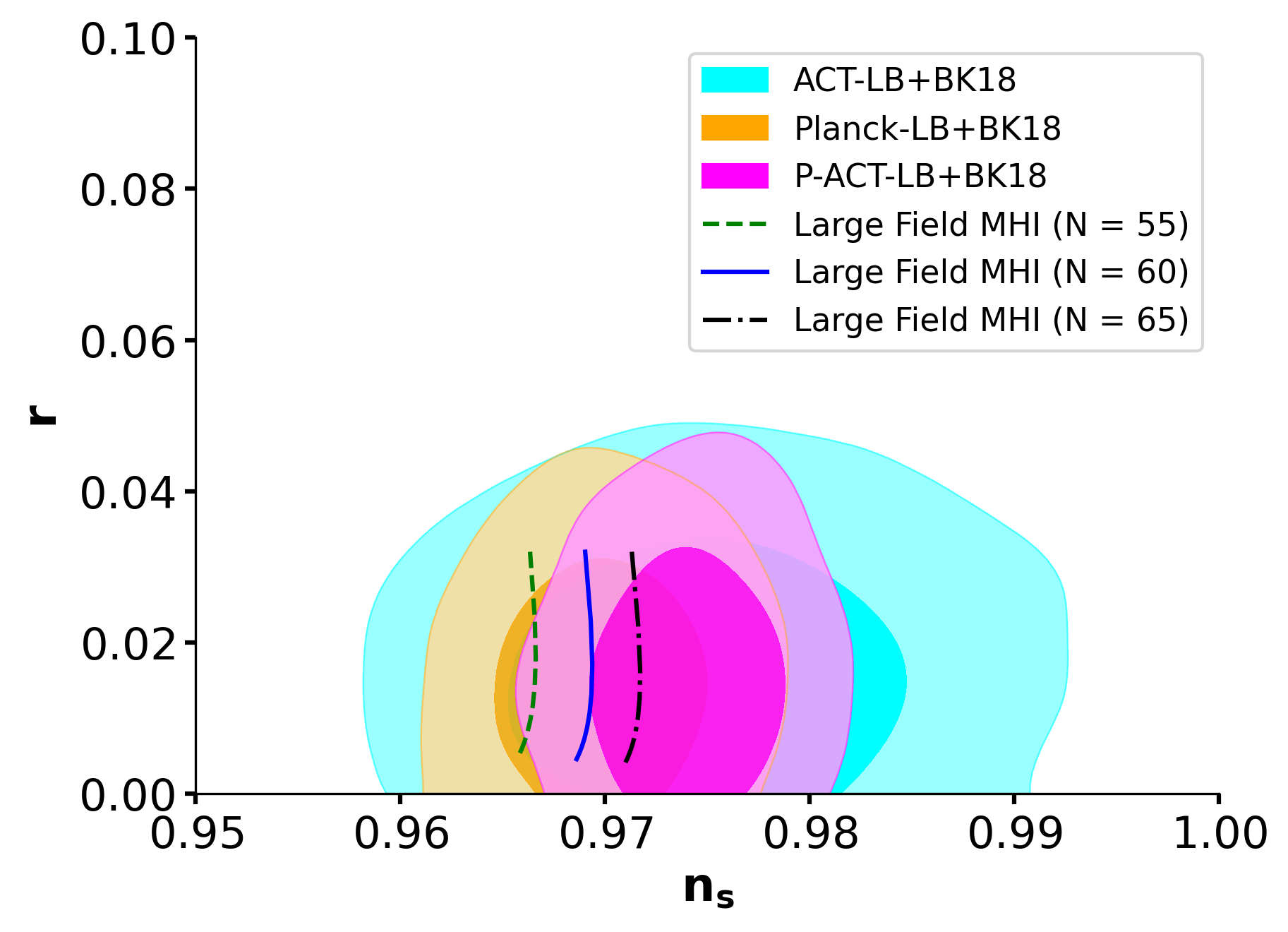}}
	\caption{\label{fig_rns-pact_large}Variation of the tensor-to-scalar ratio, $r$, with the scalar spectral index, $n_{_S}$, for three different values of e-foldings, $N = 55,\ 60,\ 65$. Black, blue, and green dashed lines correspond to predictions from the MHI (Large Field Sector) model for varying values of the model parameter and for $N = 55,\ 60,\ 65$, respectively. The marginalized $68\%$ and $95\%$ confidence regions in the plane of $r-n_{_S}$ from the joint analysis of ACT, Planck-2018 joint with BK18 and DESI-Y1 data \cite{calabrese2025atacama}  where the constraint on $r$ is obtained from BK18 data~\cite{ade2018constraints} and the constraint on $n_{_S}$ is driven by the combination of Planck, ACT DR6, and DESI-Y1 data.}
\end{figure}

\begin{figure}%[htb]
	\centerline{\includegraphics[width=15.cm, height=9.cm]{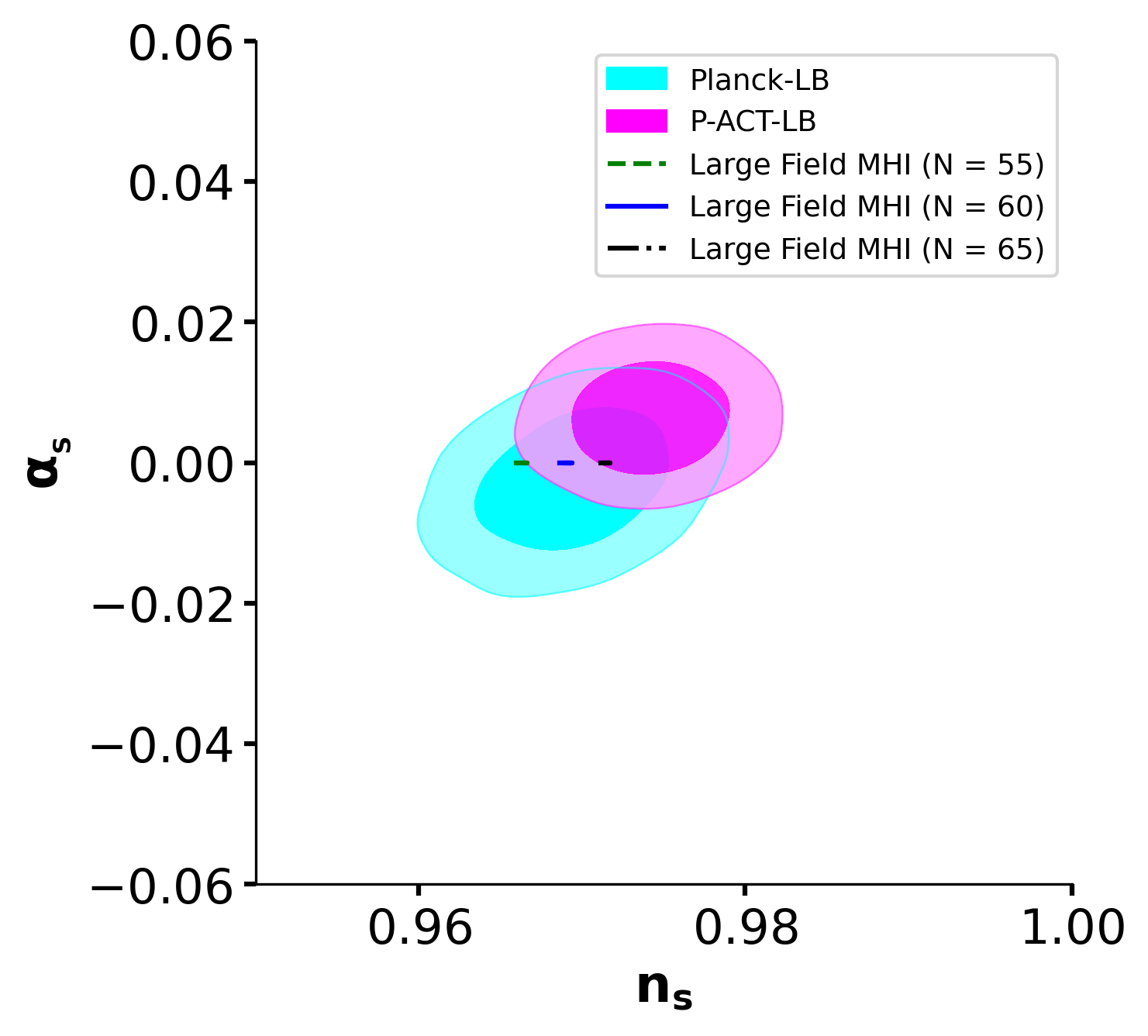}}
	\caption{\label{fig_running_large}Running of scalar spectral index, $\alpha_{_S}$, with  $n_{_S}$, in large-field MHI for three different values of e-foldings,  $N=55,\ 60, \ 65$ in black, blue and green dashed lines respectively and for varying values of the model parameter, $\alpha$. Left Panel:  Marginalized $68\%$ and $95\%$ confidence regions in the plane of $\alpha_{_S}-n_{_S}$ from the Planck-2018 data \cite{planck2015inf, ade2018constraints, ade2021improved} joint with BK18  \cite{tristram2022improved}. Right Panel:   $68\%$ and $95\%$ confidence contours in the $\alpha_{_S}-n_{_S}$ plane. The constraints are driven by the  combination of Planck, ACT-DR6 and DESI Y1 data \cite{calabrese2025atacama}}
\end{figure}

Now, in \fig{fig_rns-pact_large} we have shown predictions from large-field MHI in the plane of $r-n_{_S}$. The contours represent marginalized $68\%$ and $95\%$ confidence region from the combined analysis of ACT, Planck-2018 joint with BK18 and DESI-Y1 data \cite{calabrese2025atacama}. Although in order to explain higher observed value of $n_{_S}$, large-field MHI requires increased value for the number of e-foldings. In \fig{fig_running_large} we have compared our model prediction for running of the spectral index with the two different combination of data set \cite{calabrese2025atacama}. Though large-field MHI predicts very small amount of scalar running consistent with zero, but falls well within the observationally allowed region as predicted by PACT-LB. 
 Therefore, from the above figures \fig{fig_rns-planck_large}, \fig{fig_running-planck_large},  \fig{fig_rns-pact_large} and \fig{fig_running_large}, we may conclude that the predictions from  large-field MHI model are consistent with both Planck-2018 and P-ACT-LB results over a wide range of model parameter and for different values of $N$.

%%%%%%%%%%%%%%%%%%%%%%%%%%%%%%%%%%%%%%%%%%%%%%%%%%%%%%%%%%%%%%%%%%%%%%%%%%%%%%%%%%%%%%%
\subsection{Small Field MHI} 
\begin{figure}
	\centering
	\begin{subfigure}{.5\textwidth}
		\centering
		\includegraphics[width=.95\linewidth,height=6cm]{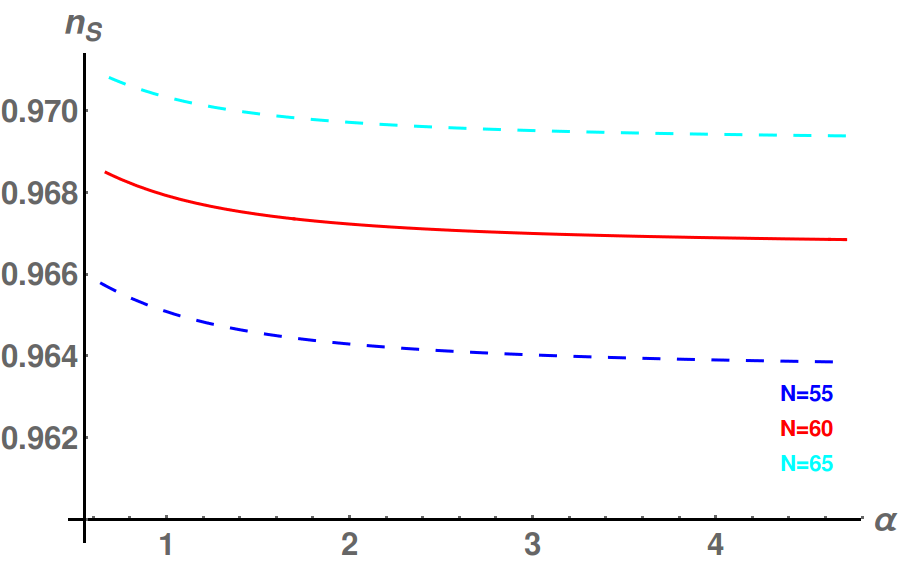}
		\caption{Scalar spectral index}
		%\label{fig:sub1}
	\end{subfigure}%
	\begin{subfigure}{.5\textwidth}
		\centering
		\includegraphics[width=.95\linewidth,height=6cm]{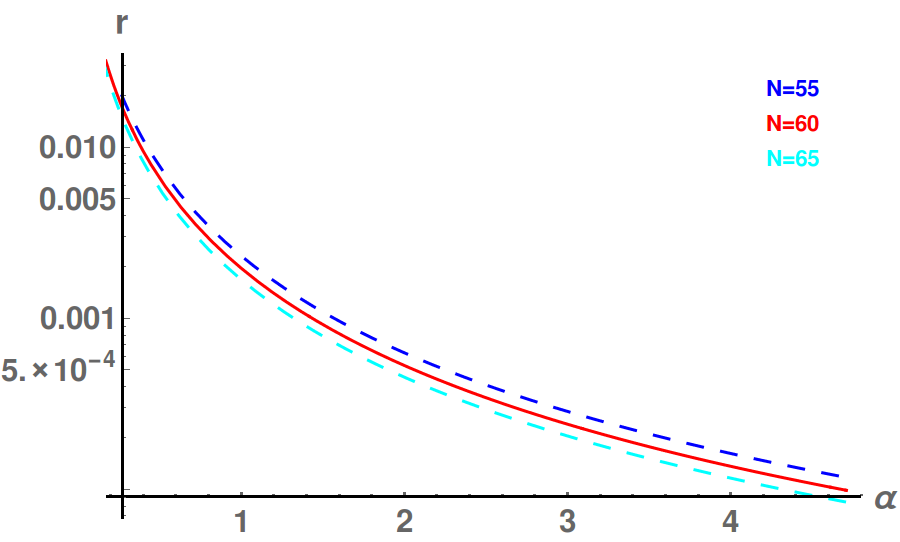}
		\caption{Tensor-to-scalar ratio}
		%\label{fig:sub2}
	\end{subfigure}
	\caption{Left Panel: The plot of scalar spectral index with the model parameter, $\alpha$, for three different values of e-foldings in small-field sector of MHI. Right Panel: Plot of tensor-to-scalar ratio with the model parameter for $N=55,\ 60, \ 65$ for small-field sector of MHI.}
	\label{fig_rns-small}
\end{figure}

We repeat the same analysis as above but now for the small-field MHI, characterized by $\Delta\phi<  \ {\rm m_P}$, and due to the lower field excursion required to unfold inflationary paradigm, this sector of MHI produce lower amount of primordial gravitational waves. In \fig{fig_rns-small}, we have shown variation of $n_{_S}$ and $r$ with the  of  model parameter. Here also, the spectral index remains almost invariant with the model parameter, but tensor-to-scalar ratio varies inversely with $\alpha$. 
The small-field branch of the MHI model can produce tensor-to-scalar ratios as low as $r\sim\mathcal{O}(10^{-4})$ while remaining consistent with current observational constraints on the scalar spectral index, and, in principle, spans the entire range of $r$ values accessible to present observations.

\begin{figure}[htb]
	\centerline{\includegraphics[width=15.cm, height=8cm]{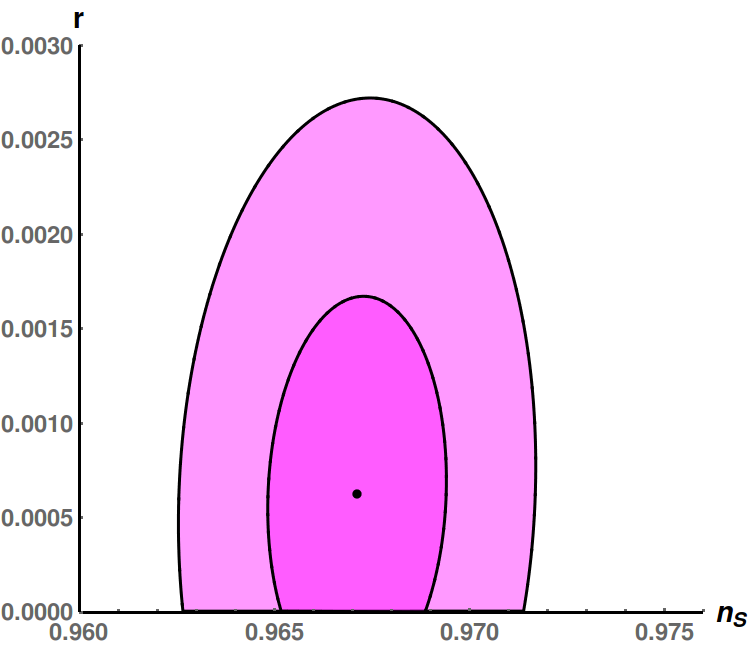}}
	\caption{\label{fig_rns-small-model}
		The $1\sigma$ and $2\sigma$ confidence regions in the $r$--$n_{_S}$ plane predicted by the small-field sector of the Mutated Hilltop Inflation (MHI) model. The contours are obtained by varying the model parameters $\alpha$ and $N$ within the parameter space corresponding to the small-field regime. The black point denotes the mean prediction of the model, while the shaded magenta regions illustrate the theoretical spread arising from variations in $\alpha$ and $N$.
	}
	
\end{figure}

As shown in \fig{fig_rns-small-model}, the 1-$\sigma$ and 2-$\sigma$ confidence regions in the $r$--$n_{_S}$ plane, corresponding to the small-field sector of MHI model, are obtained by varying the model parameters $\alpha$  within the parameter space associated with the small-field regime along with $N$ between $55$ and $65$. The mean prediction of the model is indicated by the black point, while the shaded magenta regions represent the theoretical spread resulting from variations in $\alpha$ and $N$. This figure indicates that small-field MHI can explain observations having tensor-to-scalar ratio up to $r\sim\mathcal{O}(10^{-3})$. 

\begin{figure}[htb]
	\centerline{\includegraphics[width=15.cm, height=8cm]{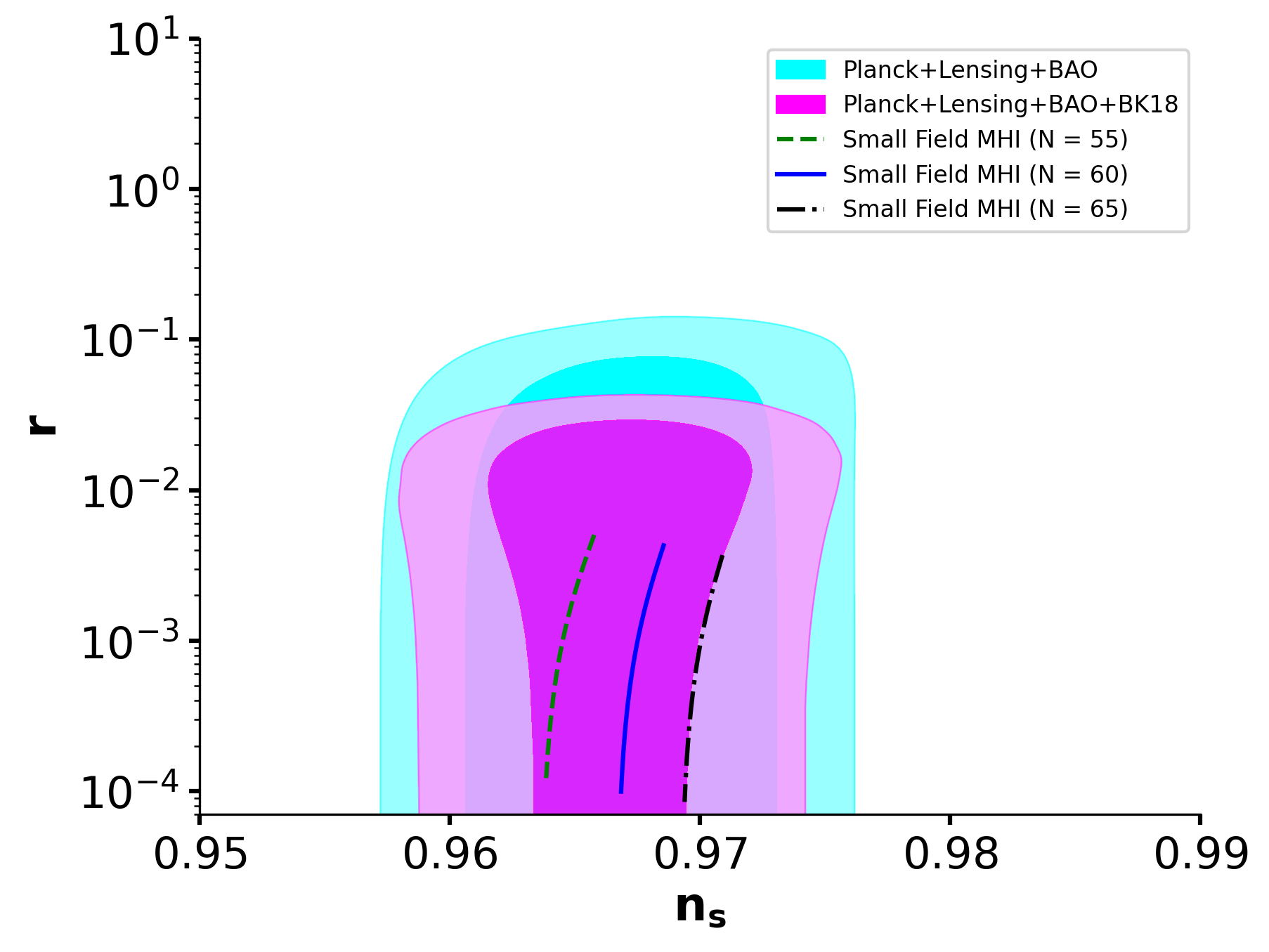}}
	\caption{\label{fig_rns_small} MHI (Small Field Sector) model prediction of the tensor-to-scalar ratio versus the scalar spectral index in the  $r-n_{_S}$ plane for three different values of e-foldings, $N = 55,\ 60,\ 65$ in green, blue, and black lines respectively. Shaded area represents marginalized $68\%$ and $95\%$ confidence contours in the $r-n_{_S}$ plane from the Planck-2018 data \cite{aghanim2020planck, ade2021improved} with and without BK18 \cite{tristram2022improved}, where the constraint on $r$ is derived from BK18 data~\cite{ade2018constraints}, while the constraint on $n_{_S}$ is driven by  Planck-2018 data.}
\end{figure}

\begin{figure}%[htb]
	\centerline{\includegraphics[width=15.cm, height=8.cm]{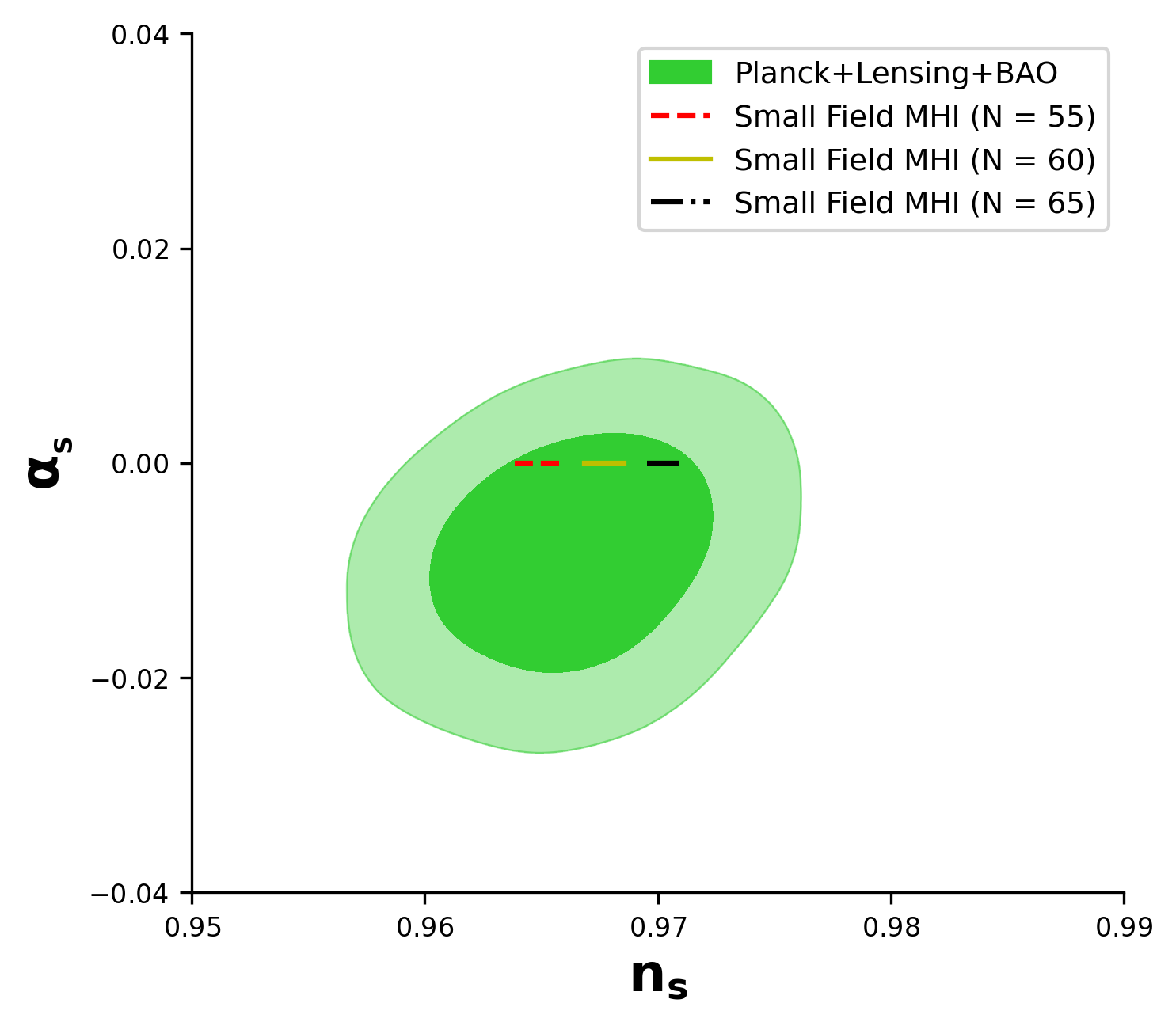}}
	\caption{\label{fig_small_running} MHI (Small Field Sector) model Prediction in the $\alpha_{_S}-n_{_S}$ plane for three different values of e-foldings, $N = 55,\ 60,\ 65$ in green, blue, and black lines. Shaded area represents marginalized $68\%$ and $95\%$ confidence contours in the $\alpha_{_S}-n_{_S}$ plane, constraints  are  driven by  Planck-2018 data \cite{aghanim2020planck, ade2021improved}.}
\end{figure}

In \fig{fig_rns_small} we have shown variation of the tensor-to-scalar ratio with spectral index for the small-field MHI. Shaded contours represent marginalized  $68\%$ and $95\%$ confidence level in the $r-n_{_S}$ plane, where the constraints are driven by the combination of Planck-2018 and BK18 data. Whereas  \fig{fig_small_running} depicts prediction from MHI (Small Field Sector) in the $\alpha_{_S}-n_{_S}$ plane,  constraints being driven by Planck-2018 data. In both the cases we find that small-field MHI is in excellent agreement for wide range of the model parameter. 

\begin{figure}%[htb]
	\centerline{\includegraphics[width=15.cm, height=9.cm]{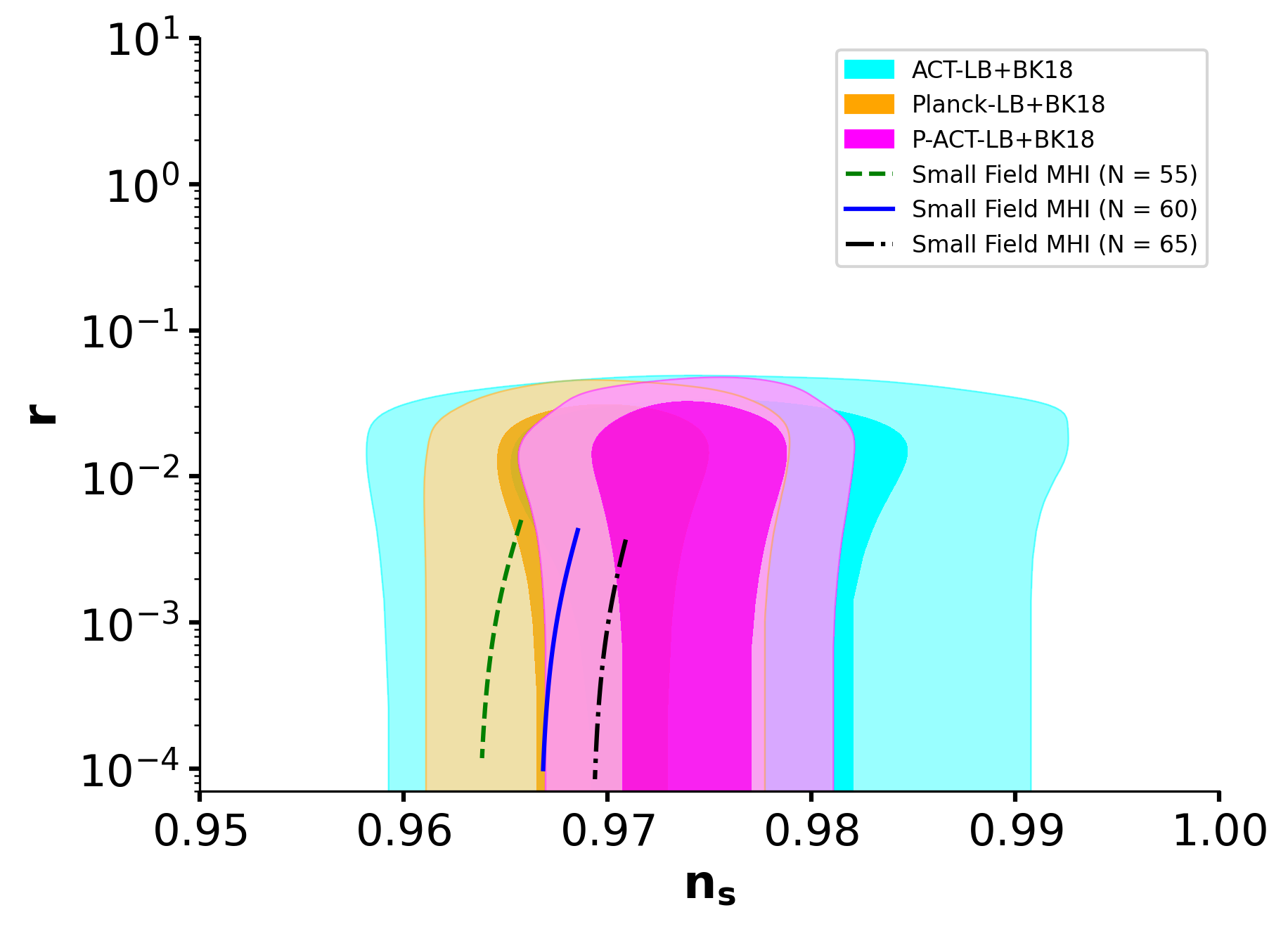}}
	\caption{\label{fig_rns_small-pact}Prediction of the tensor-to-scalar ratio versus the scalar spectral index in their plane for three different values of e-foldings, $N = 55,\ 60,\ 65$ in green, blue and black lines respectively, from the small-field MHI. Shaded area representing marginalized $68\%$ and $95\%$ confidence contours in the $r-n_{_S}$ plane. The constraint on $r$ is driven by from BK18 data~\cite{ade2018constraints}, while the constraint on $n_{_S}$ is obtained from the combined analysis of ACT, Planck-2018 and DESI-Y1 data \cite{calabrese2025atacama}.}
\end{figure}

\begin{figure}%[htb]
	\centerline{\includegraphics[width=15.cm, height=9.cm]{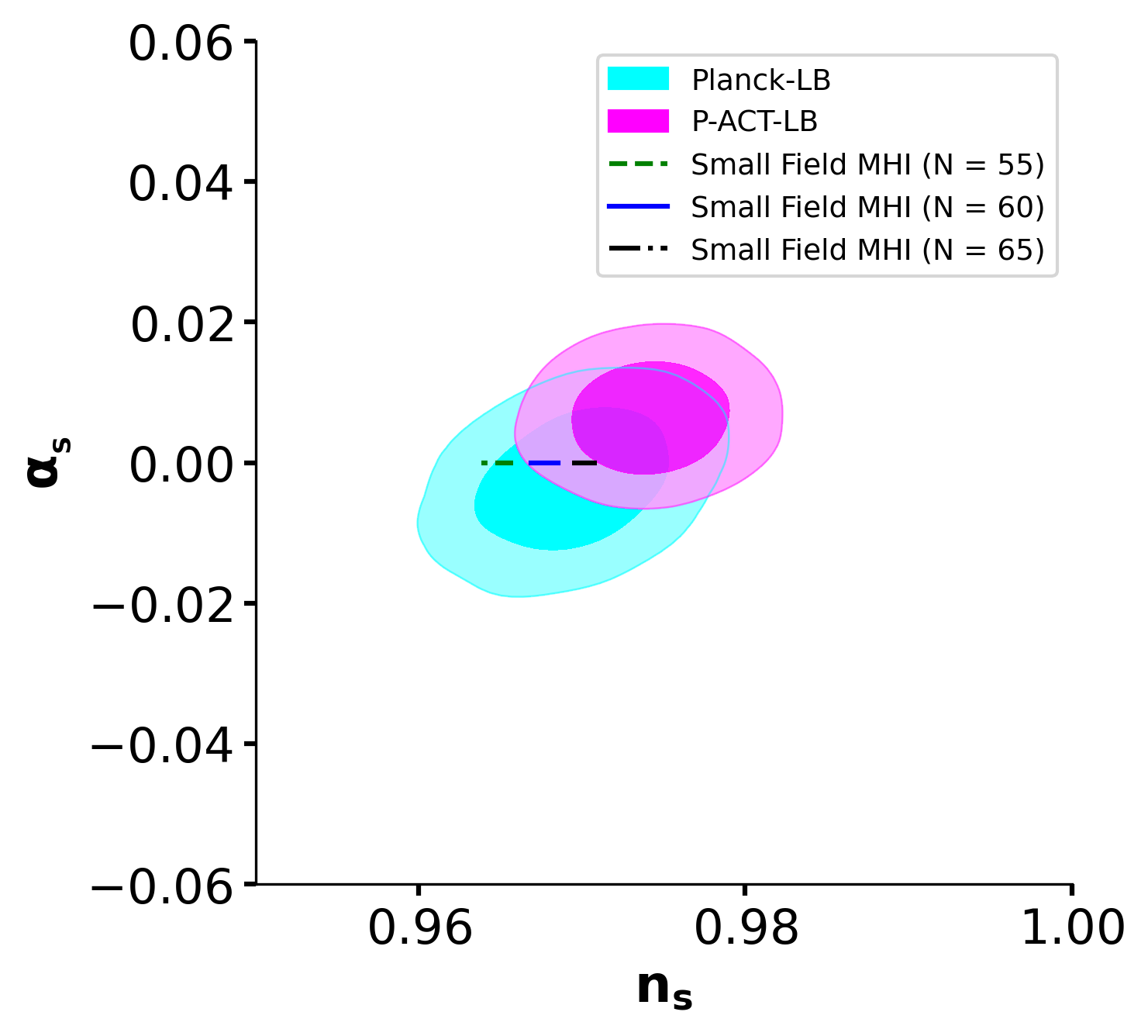}}
	\caption{\label{fig_small_running-pact} Running of scalar spectral index, $\alpha_{_S}$, with  $n_{_S}$, in small-field MHI for three different values of e-foldings,  $N=55,\ 60, \ 65$ in black, blue and green dashed lines respectively and for varying values of the model parameter, $\alpha$. The marginalized $68\%$ and $95\%$ confidence contours in the $\alpha_{_S}-n_{_S}$ plane is constrained by the combined analysis of ACT, Planck-2018 and DESI-Y1 data.}
\end{figure}

In \fig{fig_rns_small-pact} we have shown variation of the tensor-to-scalar ratio versus spectral index for the small-field sector of MHI. Shaded contours represent marginalized  $68\%$ and $95\%$ confidence level in the $r-n_{_S}$ plane, where the constraints are driven by the combination of ACT, Planck-2018 joint with BK18 and DESI-Y1 data. Here also we may notice that, small-field MHI demands elevated number of e-foldings to account for the observed higher value of scalar spectral index by ACT. 
Whereas  \fig{fig_small_running-pact} depicts prediction from small-field MHI in the $\alpha_{_S}-n_{_S}$ plane,  constraints being driven jointly by ACT, Planck-2018 and DESI-Y1 data. Here again we see that predictions from small-field MHI lie within allowed region as predicted by the latest ACT-DR6 results.  

%%%%%%%%%%%%%%%%%%%%%%%%%%%%%%%%%%%%%%%%%%%%%%%%%%%%%%%%%%%%%%%%%%%%%%%%%%%%%%%%%%%%%%%
\section{MHI and FUTURE CMB Experiments}
LiteBIRD is a space mission for primordial cosmology and fundamental physics,  promising to detect  
tensor-to-scalar ratio with uncertainty of $\sigma( r)\sim0.001$ \cite{litebird2023probing, ghigna2024litebird}. Non-detection of CMB B-mode polarization, LiteBIRD will impose an upper bound on the amplitude of primordial gravitational waves $r<0.002$. 
While, Simons Observatory \cite{SO2019}, a forthcoming ground-based CMB experiment designed to measure temperature and polarization anisotropies with unprecedented sensitivity and angular resolution, is optimistic about  touching the sensitivity of $\sigma( r)\sim0.003$. Here, we assess the consistency of the mutated hilltop inflation  model predictions with the forecasted sensitivities of upcoming CMB experiments, LiteBIRD and the Simons Observatory.

\begin{figure}
	\centering
	\begin{subfigure}{.5\textwidth}
		\centering
		\includegraphics[width=.95\linewidth,height=6cm]{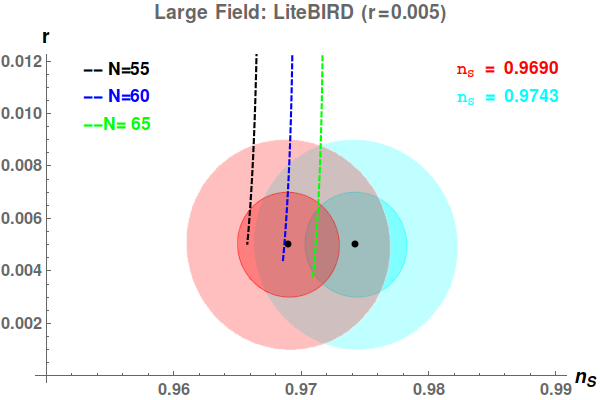}
		\caption{LiteBIRD}
		%\label{fig:sub1}
	\end{subfigure}%
	\begin{subfigure}{.5\textwidth}
		\centering
		\includegraphics[width=.95\linewidth,height=6cm]{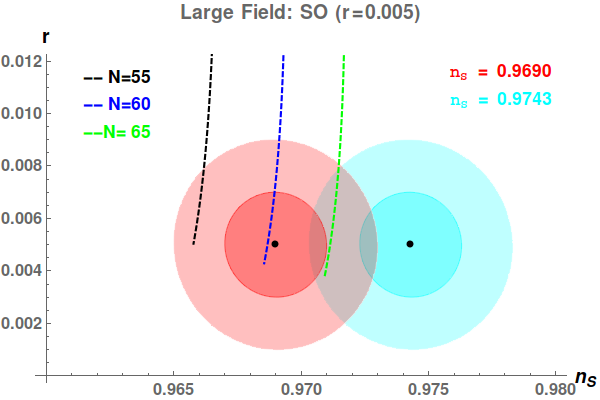}
		\caption{SO}
		%\label{fig:sub2}
	\end{subfigure}
	\caption{Tensor-to-scalar ratio, $r$, with the scalar spectral index, $n_{_S}$, for three different values of e-foldings,  $N=55,\ 60, \ 65$ in black, blue, and green dashed lines respectively indicating the prediction of $r$ and $n_{_S}$ from large-field  MHI and for different values of the model parameter. \textbf{Left Panel:}  LiteBIRD constraints from a fiducial model with $r=0.01,\ 0.005$ and $n_{_S}=0.9743$. Forecasted marginalized $68\%$ and $95\%$ confidence contours in the $r-n_{_S}$ plane  expecting sensitivity of the LiteBIRD experiment. \textbf{Right Panel:} Forecasted marginalized $68\%$ and $95\%$ confidence contours from a fiducial model with $r=0.01,\ 0.005$ and $n_{_S}=0.9743$.  in the $r-n_{_S}$ plane.  The constraints reflect  expected sensitivity of the SO experiment. The black dots at the centre of each contour represent fiducial model.}
	\label{fig_rns_LB-S4-Large}
\end{figure}

\begin{figure}
	\centering
	\begin{subfigure}{.5\textwidth}
		\centering
		\includegraphics[width=.95\linewidth,height=6cm]{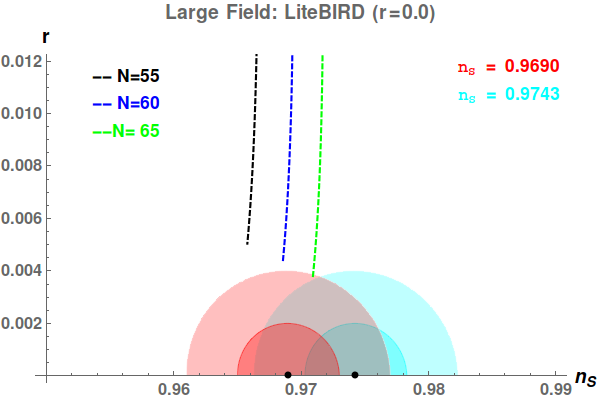}
		\caption{LiteBIRD}
		%\label{fig:sub1}
	\end{subfigure}%
	\begin{subfigure}{.5\textwidth}
		\centering
		\includegraphics[width=.95\linewidth,height=6cm]{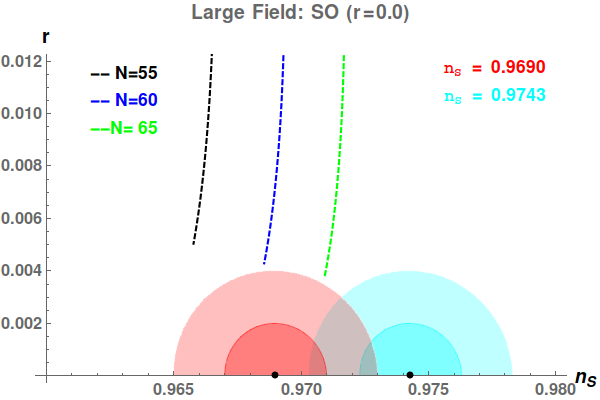}
		\caption{CMB-S4}
		%\label{fig:sub2}
	\end{subfigure}
	\caption{Forecasted marginalized $68\%$ and $95\%$ confidence contours in the $r-n_{_S}$ plane for a fiducial model with $r=0.0$ and $n_{_S}=0.9651,\ 0.9743$. Black, blue, and green dashed lines indicate the predictions from large-field MHI for  $N=55,\ 60,\ 65$ respectively. \textbf{Left Panel:} Statistical uncertainties on $r$ and $n_{_S}$, assuming Gaussian likelihoods as expected from the sensitivity of LiteBIRD mission.  \textbf{Right Panel:} The constraint  are driven by the  expected sensitivity of the futuristic SO mission. The black dots at the centre of each contour represent fiducial model.}
	\label{fig_rns_LB-S4-Large-r0}
\end{figure}

In \fig{fig_rns_LB-S4-Large} we have depicted tensor-to-scalar ratio with scalar spectral index as expected from large-field MHI in the $r-n_{_S}$ plane for three different values of e-foldings and varying model parameter. The figure portrays forecasted marginalized joint $68\%$ and $95\%$ confidence contours in the $r-n_{_S}$ plane anticipating  sensitivity of the  futuristic CMB missions LiteBIRD and SO,  in the left and right panel respectively,  for a fiducial model with $r=0.005$ and $n_{_S}=0.9690, \ 0.9743$ \cite{tristram2024, calabrese2025atacama}. A careful consideration of the plot shows that large-field MHI provides an excellent match with those futuristic missions. However non-detection of primordial gravitational waves by  LiteBIRD or SO does not favour large-field MHI as depicted in \fig{fig_rns_LB-S4-Large-r0}. 

\begin{figure}
	\centering
	\begin{subfigure}{.5\textwidth}
		\centering
		\includegraphics[width=.95\linewidth,height=6cm]{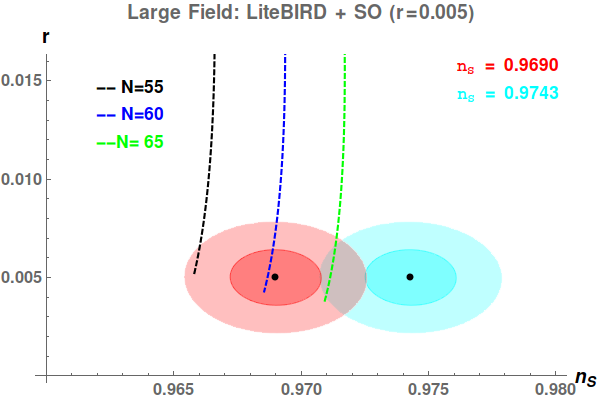}
		\caption{$r=0.005$}
		%\label{fig:sub1}
	\end{subfigure}%
	\begin{subfigure}{.5\textwidth}
		\centering
		\includegraphics[width=.95\linewidth,height=6cm]{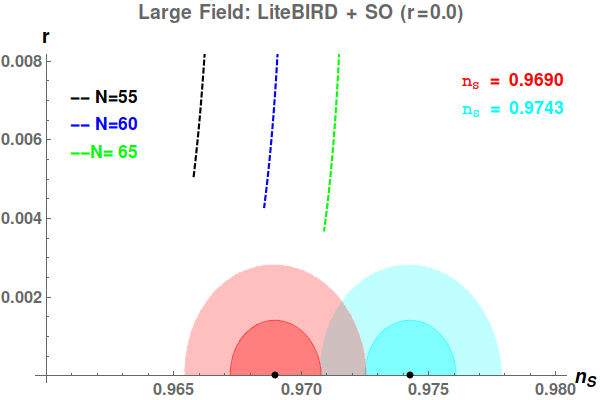}
		\caption{$r=0.0$}
		%\label{fig:sub2}
	\end{subfigure}
	\caption{Forecasted marginalized $68\%$ and $95\%$ confidence contours in the $r-n_{_S}$ plane for  fiducial values $r=0.0,\ 0.005$ and $n_{_S}=0.9690,\ 0.9743$. The constraint on $r$ and $n_{_S}$ are driven by the  expected sensitivity of the joint analysis of futuristic space mission LiteBIRD and ground based SO mission. Black, blue and green dashed lines represent expectation  from large-field  MHI in the $r-n_{_S}$ plane for  $N=55,\ 60,\ 65$ respectively. The black dots at the centre of each contour represent fiducial model. \textbf{ Left Panel:} Here we have considered $r=0.005$ and  $n_{_S}=0.9690,\ 0.9743$ from Planck-2024 result  \cite{tristram2024} and P-ACT-LB analysis \cite{calabrese2025atacama} respectively . \textbf{Right Panel:} Here we have used  $r=0.0$ and  $n_{_S}=0.9690,\ 0.9743$ from Planck-2024 result  \cite{tristram2024} and P-ACT-LB analysis \cite{calabrese2025atacama} respectively.}
	\label{fig_rns_LBSO-Large}
\end{figure}

In \fig{fig_rns_LBSO-Large} we have presented forecasts of joint constraints on tensor-to-scalar ratio and scalar spectral index assuming fiducial values $r=0.005$ (in the Left Panel) and $r=0.0$ (in the Right Panel) along with $n_{_S}=0.9690, \ 0.9743$. Constraints on $n_{_S}$ and $r$ are derived from expected sensitivity of  joint analysis of LiteBIRD and SO.  We see that large-field MHI is consistent with the joint analysis for fiducial values $r=0.005$ and $n_{_S}=0.9690, \ 0.9743$. However non-detection of primordial gravitational waves by LiteBIRD and SO will rule out large-field MHI. 

\begin{figure}
	\centering
	\begin{subfigure}{.5\textwidth}
		\centering
		\includegraphics[width=.95\linewidth,height=6cm]{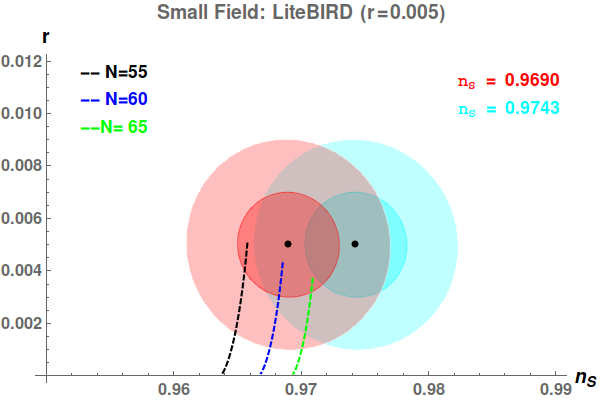}
		\caption{LiteBIRD}
		%\label{fig:sub1}
	\end{subfigure}%
	\begin{subfigure}{.5\textwidth}
		\centering
		\includegraphics[width=.95\linewidth,height=6cm]{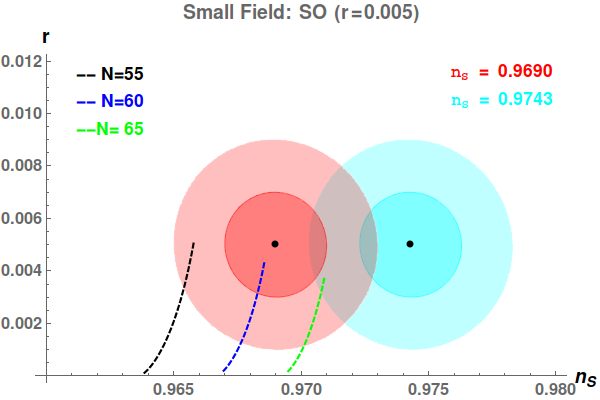}
		\caption{Simons Observatory}
		%\label{fig:sub2}
	\end{subfigure}
	\caption{Tensor-to-scalar ratio, $r$, with the scalar spectral index, $n_{_S}$, for three different values of e-foldings,  $N=55,\ 60, \ 65$ in black, blue, and green dashed lines respectively indicating the prediction of $r$ and $n_{_S}$ from small-field  MHI and varying model parameter, $\alpha$. \textbf{Left Panel:}  LiteBIRD constraints from a fiducial model with $r=0.005$ and $n_{_S}=0.9743$. Forecasted marginalized $68\%$ and $95\%$ confidence contours in the $r-n_{_S}$ plane  expecting sensitivity of the LiteBIRD experiment. \textbf{Right Panel:} Forecasted marginalized $68\%$ and $95\%$ confidence contours from a fiducial model with $r=0.005$ and $n_{_S}=0.9743$.  in the $r-n_{_S}$ plane.  The constraints reflect  expected sensitivity of the SO experiment.The black dots at the centre of each contour represent fiducial model.}
	\label{fig_rns_LB-SO-Small}
\end{figure}

%%%%%%%%%%%%%%%%%%%%%%%%%%SMALL FIELD MHI
%%%%%%%%%%%%%%%%%%%%%%%%%%%%%%%%%%%%%%%%%%%%%%%%%%%%%%%%%%%%%%%5
\fig{fig_rns_LB-SO-Small} shows  expected variation of tensor-to-scalar ratio with scalar spectral index from small-field sector of  MHI in the $r-n_{_S}$ plane for three different values of $N$ and varying $\alpha$. The figure represents forecasted marginalized joint $68\%$ and $95\%$ confidence contours in the $r-n_{_S}$ plane anticipating  sensitivity of the  futuristic CMB missions LiteBIRD and SO,  in the left and right panel respectively,  for a fiducial model with $r=0.005$ and $n_{_S}=0.9690, \ 0.9743$ \cite{tristram2024, calabrese2025atacama}. A careful consideration of the plot shows small-field MHI provides an excellent match with those futuristic missions. 

\begin{figure}
	\centering
	\begin{subfigure}{.5\textwidth}
		\centering
		\includegraphics[width=.95\linewidth,height=6cm]{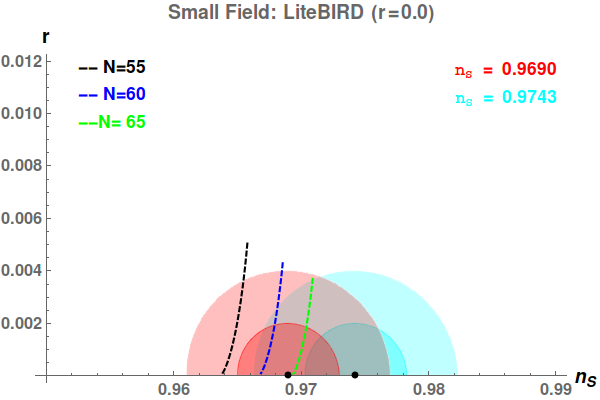}
		\caption{LiteBIRD}
		%\label{fig:sub1}
	\end{subfigure}%
	\begin{subfigure}{.5\textwidth}
		\centering
		\includegraphics[width=.95\linewidth,height=6cm]{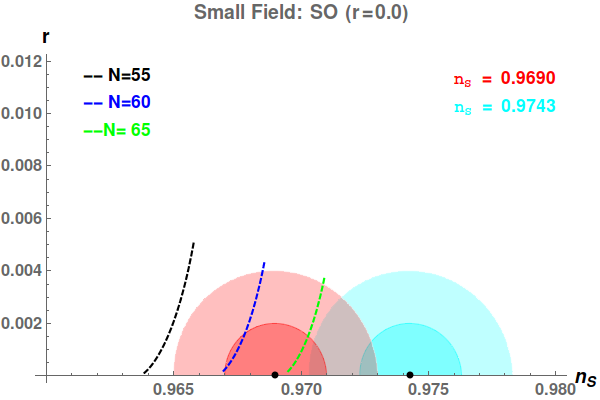}
		\caption{Simons Observatory}
		%\label{fig:sub2}
	\end{subfigure}
	\caption{Prediction of the small-field MHI in the $r-n_{_S}$ plane for three different values of e-foldings,  $N=55,\ 60, \ 65$ in black, blue, and green dashed lines respectively. \textbf{Left Panel:}  LiteBIRD forecasted marginalized $68\%$ and $95\%$ confidence contours in the $r-n_{_S}$ plane for a fiducial model with $r=0.0$ and $n_{_S}=0.9690,\ 0.9743$.  \textbf{Right Panel:} Forecasted marginalized $68\%$ and $95\%$ confidence contours from a fiducial model with $r=0.0$ and $n_{_S}=0.9690,\ 0.9743$,  in the $r-n_{_S}$ plane.  The constraints reflect  expected sensitivity of the SO experiment.  In each panel the black dots at the centre of contours mark the corresponding fiducial models.}
	\label{fig_rns_LB-SO-Small-r0}
\end{figure}

In \fig{fig_rns_LB-SO-Small-r0} we have demonstrated predictions from  small-field MHI in the $r-n_{_S}$ plane for different values of $N$ and $\alpha$. The shaded region corresponds to forecasted marginalized $68\%$ and $95\%$ confidence region in the $r-n_{S}$ plane, for the expected sensitivity of the LiteBIRD and SO experiments,  for a fiducial model with $r = 0.0$ and $n_{_S} = 0.969,\ 0.9743$. This result is particularly noteworthy, as it demonstrates that even a non-detection of primordial gravitational waves by future CMB missions may not be sufficient to rule out the small-field MHI scenario.

\begin{figure}
	\centering
	\begin{subfigure}{.5\textwidth}
		\centering
		\includegraphics[width=.95\linewidth,height=6cm]{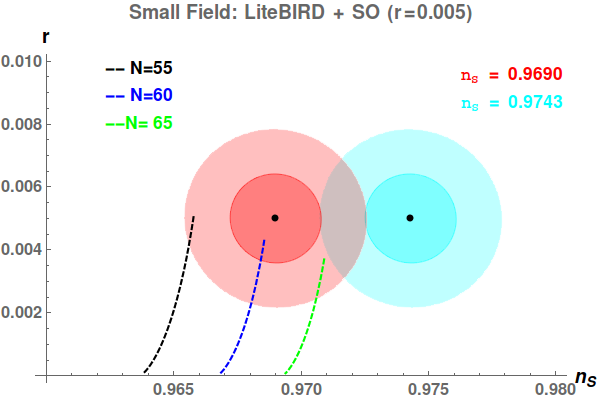}
		\caption{$r=0.005$}
		%\label{fig:sub1}
	\end{subfigure}%
	\begin{subfigure}{.5\textwidth}
		\centering
		\includegraphics[width=.95\linewidth,height=6cm]{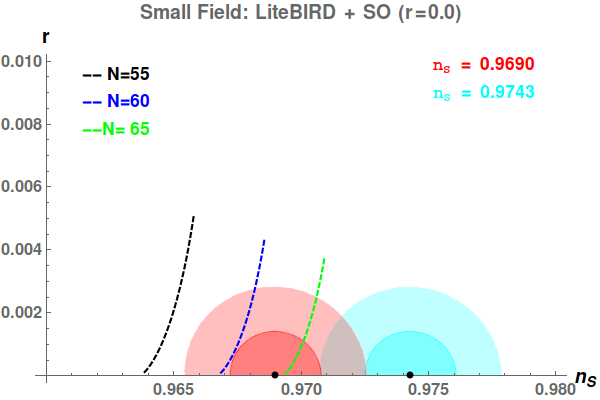}
		\caption{$r=0.0$}
		%\label{fig:sub2}
	\end{subfigure}
	\caption{Forecasted marginalized $68\%$ and $95\%$ confidence contours in the $r-n_{_S}$ plane for fiducial values $r=0.005$ and $n_{_S}=0.9690,\ 0.9743$. The constraints on $r$ and $n_{_S}$ are driven by the expected sensitivity of a joint analysis combining the futuristic space mission LiteBIRD with the ground-based SO experiment. Black, blue and green dashed lines show the predictions of the small-field MHI for $N=55,\,60,\,65$, respectively. Black dots at the centres of the contours mark the fiducial models. 
		\textbf{Left panel:} $r=0.005$ with $n_{_S}=0.9651$ (Planck 2018~\cite{ade2021improved}) and $n_{_S}=0.9743$ (P-ACT-LB analysis~\cite{calabrese2025atacama}). 
		\textbf{Right panel:} $r=0.0$ with the same $n_{_S}$ values from Planck 2018 and P-ACT-LB.}
	\label{fig_rns_LBSO-Small}
\end{figure}
In \fig{fig_rns_LBSO-Small}, we illustrate the forecasted constraints in the $r$–$n_{_S}$ plane obtained from a idealistic joint analysis of the future space-based LiteBIRD mission and the ground-based SO experiment. The forecasted marginalized $68\%$ and $95\%$ confidence contours are shown for fiducial models characterized by $r = 0.005$ and $n{_S} = 0.9690,\ 0.9743$. The corresponding predictions of the small-field MHI model are depicted by the black, blue and green dashed curves for $N = 55,\ 60,\ 65$, respectively. As evident from the figure, the combined sensitivity of LiteBIRD and SO is expected to significantly improve constraints on the tensor-to-scalar ratio and the scalar spectral index compared to current limits. Nevertheless, even in the case of a null detection of primordial gravitational waves ($r=0$), the small-field MHI scenario remains consistent with observational forecasts, indicating that future non-detections alone may not be sufficient to rule out this class of inflationary models.

%%%%%%%%%%%%%%%%%%%%%%%%%%%%%%%%%%%%%%%%%%%%%%%%%%%%%%%%%%%%%%%%%%%%%%%%%%%%%%%%%%%%%%%%%%%%%%%%%%%%%%%%%%%%%%%%%%%%%%%%%%%%%%%%%%%%%%%%%%%%%%%%%%%%%%%%%%%%%%%%%%%%%%%%%%%%%%%%%%%%%%%%%%%%%%%%%%%%%%%%%%%%%%%%%%%%%%%%%%%%%%%%%%%%%%%%
\section{Conclusion}\label{con}
In this short article, we have examined the advantages and limitations of the MHI model. Depending on the choice of the model parameter, the MHI model admits two distinct branches that can, in principle, probe a wide range of tensor-to-scalar ratios, making it challenging to rule out even with future CMB missions aimed at detecting primordial gravitational waves. A distinctive feature of the MHI model is that it predicts scalar perturbations with a spectral index $n_{_S}$ that remains independent of the model parameter.

Future joint analyses combining LiteBIRD’s high-precision measurements of large-scale CMB polarization with the small-scale sensitivity of the Simons Observatory are expected to significantly tighten the constraints on the tensor-to-scalar ratio. Such advances will provide a powerful means to further probe the inflationary parameter space and evaluate the viability of  MHI model relative to other inflationary scenarios. Overall, the MHI model stands out as a observationally resilient framework for describing the dynamics of the early Universe.

The  flexibility of  MHI model, combined with its strong compatibility with current observational limits, makes it a compelling candidate for describing the early Universe. With forthcoming CMB missions such as LiteBIRD and ground-based  SO, the MHI  offers a promising avenue to explore the physics of the primordial epoch. One of the most fascinating aspects of the small-field branch of the MHI model is that it remains consistent with observations even in the absence of a primordial gravitational-wave detection by LiteBIRD and SO. 
We have also found that spectral index as predicted by MHI model is  almost independent of $\alpha$, but strongly depends on the number of e-foldings. Consequently, accommodating the relatively high observational value of the scalar spectral index may necessitate a higher e-folding number within the MHI framework.

Overall, observational resilience, and predictive flexibility make the MHI model a compelling candidate for describing the inflationary epoch. It provides a framework capable of accommodating both the detection and non-detection of primordial gravitational waves, making it one of the most versatile models for understanding the dynamics of the early Universe.
%%%%%%%%%%%%%%%%%%%%%%%%%%%%%%%%%%%%%%%%%%%%%%%%%%%%%%%%%%%%%%%%%%%%%%%%%%%%

%\bibliographystyle{plain}
\bibliographystyle{unsrt}
\bibliography{references}

\end{document}